\documentclass[aps,prd,onecolumn,11pt,groupedaddress,nofootinbib,showpacs]{revtex4}
\usepackage{amsmath,amssymb}
\usepackage[dvips]{graphicx}
\begin{document}

\title{Replica symmetry breaking
% of the Ising spin glass
in and around
six dimensions}

\author{G.~Parisi}
\email{giorgio.parisi@roma1.infn.it}
\affiliation{Dipartimento di Fisica, Universit\`a di Roma ``La Sapienza'', 
P.le Aldo Moro 2, I-00185 Roma, Italy \\
Statistical Mechanics and Complexity Center (SMC) - INFM - CNR, Italy
}
\author{T.~Temesv\'ari}
\email{temtam@helios.elte.hu}
\affiliation{
Research Group for Theoretical Physics of the Hungarian Academy of Sciences,
E\"otv\"os University, P\'azm\'any P\'eter s\'et\'any 1/A,
H-1117 Budapest, Hungary}

\date{\today}

\begin{abstract}
Two, replica symmetry breaking specific, quantities of the Ising spin glass
--- the breakpoint $x_1$ of the order parameter
function and the Almeida-Thouless line --- are calculated in six dimensions (the upper critical
dimension of the replicated field theory used), and also below and above it. The results
comfirm that replica symmetry breaking does exist below $d=6$, and also the tendency of its escalation
for decreasing dimension continues. As a new feature, $x_1$ has a nonzero and universal value for
$d<6$ at criticality.
Near six dimensions we have  $x_{1c}=3\,(6-d)+O[(6-d)^2]$.
A method to expand a generic theory with replica equivalence around the replica
symmetric one is also demonstrated.
\end{abstract}


\maketitle

\section{Introduction}

Frustration in disordered systems gives rise to a complex equilibrium state with a nontrivial
breaking of ergodicity (see \cite{MePaVi} for a review and important reprints of the field).
In the mean field version of the Ising spin glass \cite{SK}, the decomposition of the Gibbs
state into ultrametrically organized pure states is (mathematically) encoded in the replica
symmetry broken (RSB) solution of the replicated system \cite{MePaVi}. This solution has
characteristics --- such as the order parameter {\em function\/} $q(x)$, and the spin glass transition
in nonzero external magnetic field along the so called Almeida-Thouless (AT) line ---
which fully distinguish it from the much simpler replica symmetric (RS) case.
This RS solution is unstable in the mean field glassy phase \cite{AT}.

From the physical point of view, RSB implies the presence of violations of nontrivial
fluctuation-dissipation relations at off-equilibrium (during aging), while the off-equilibrium
fluctuation-dissipation relations would be trivial in the RS case: in particular 
no aging of the response function is expected then, in variance with the experimental evidence in
three dimensions at zero magnetic field. It is a very important task to determine the dimensional
regime where the low temperature phase with aging response function survives. Evidently, there is no glassy
phase in the one-dimensional system, whereas there is an ample numerical evidence against
any transition in the two-dimensional case too. Generally speaking, we expect that the transitions
disappear at the corresponding lower critical dimensions, i.e\ at  $d_{SG}^{0}$ in zero magnetic field,
and at $d_{SG}^{h}$ in the presence of a magnetic field.  We cannot say a priori if these two lower
critical dimensions are the same: in the case of an Ising ferromagnet with a random magnetic field,
for instance,
it is well known that $d_{IF}^{0}=1$, whereas $d_{IF}^{h}=0$. The situation in spin glasses is quite unclear:
the different structure of the low momentum singularities in zero and nonzero magnetic field
\cite{Gaussian_propagators} suggest that $d_{SG}^{0}<d_{SG}^{h}$, while the arguments based on
domain wall energies give $d_{SG}^{0}=d_{SG}^{h}=2.5$ \cite{Franz_Parisi_Virasoro}.
The existence of a low temperature phase with aging response function should be  ultimately
decided by investigating the structure of infrared divergences in the perturbative expansion,
and by the analysis of nonperturbative contributions. This task goes by far beyond the goals of the
present paper. We aim to study in details the properties of the low temperature phase
near the critical temperature, and around the upper critical dimension (i.e.\ six)
where the critical exponents at zero magnetic field become nontrivial.  Our study also aims to correct
some recent claims on the nonexistence of a RSB phase below six dimensions that are due to an incorrect
analysis of the consequences of some renormalization group equations \cite{Moore_Bray_2011}.

The mean field Ising spin glass, at least when studied with the replica trick, can be considered
as the infinite-dimensional limit of the replica field theory representing the $d$-dimensional
short ranged model defined on a hypercubic lattice \cite{AT2008}. The study of this replica field theory
for decreasing dimensionalities seems to be a good strategy for reaching a full understanding of the
three-dimensional Ising spin glass.

This project has had by now a long history whose first period was summarized in Ref.\ \cite{beyond}.
It turns out from these studies that the RS glassy phase is notoriously unstable even
down to $d\lesssim 6$, with a persistently escalating RSB phase (see, for instance,
Fig.\ 1 of Ref.\ \cite{nucl}). A scaling picture was proposed in \cite{scaling_and_infrared} for
helping to understand one-loop calculations in the (zero external magnetic field) RSB phase.
Some of the results of this reference are reproduced and/or revised in the present paper, especially
the behaviour of the breakpoint $x_1$ of $q(x)$ around six dimensions. The AT line was first found
in Ref.\ \cite{GrMoBr83} for the range $6<d<8$, whereas it was followed up from mean field
($d=\infty$) to $d\lesssim 6$ (and also for nonzero replica number $n$) in \cite{AT2008}.

\begin{figure}
\caption{Schematic phase diagrams for a $d$-dimensional Ising spin glass in the temperature-magnetic field
plane. There is an RSB glassy phase in (a) bordered by the AT line. On the other hand, the glassy phase
is RS in (b), and lies in the zero-field subspace. Both the AT line and the zero-field glassy phase are
represented by the same generic replica symmetric field theory with massive longitudinal and massless
replicon modes.
}
\label{RSB_vs_droplet}
\vspace{20pt}
\includegraphics[scale=1]{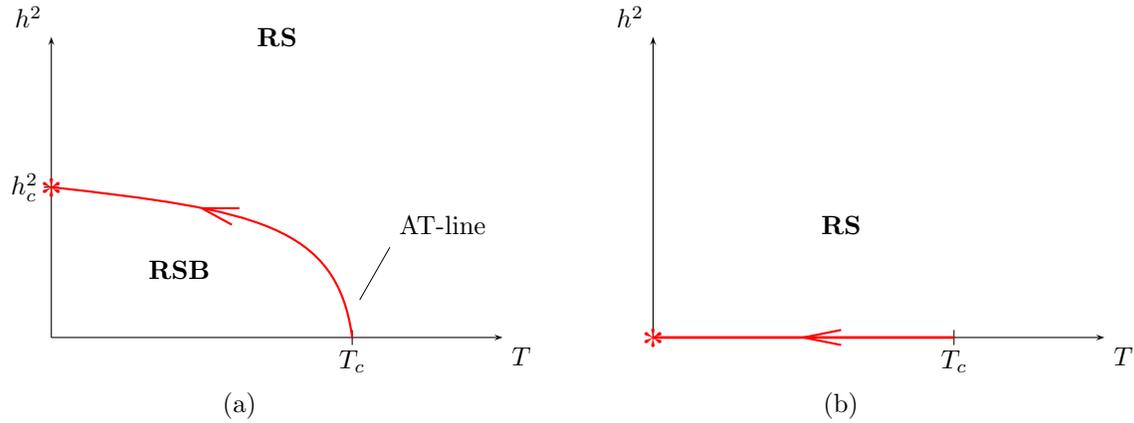}
\vspace{20pt}
\end{figure}

Nevertheless, the RS spin glass phase has remained an alternative due to the so called droplet
model \cite{FiHu86,FiHu88,BrMo86}. This theory predicts a unique Gibbs state (apart from spin inversion)
for $T<T_c$
--- that is why the replicated theory is RS --- which is massless, and the glassy phase is unstable
for any infinitesimal magnetic field, i.e.\ there is no AT line.
A schematic picture of the two scenarios on the temperature-magnetic field
plane is presented in Fig.\ \ref{RSB_vs_droplet}. The phase boundary lies
along the temperature axis in the droplet case, a zero-temperature fixed point
governing its behaviour; the analogous attractive --- and also zero-temperature ---
fixed point for the RSB scenario is shifted to a nonzero external field $h_c$. The other
end of the phase boundary is, in both cases, the zero-field critical fixed point at $T_c$.
Since the symmetry of the transition line --- namely, an RS state with nonzero order parameter $q$, which
is massless in the so called replicon sector, while massive in the longitudinal one --- is the same
(notwithstanding the fact that the AT line proceeds in nonzero magnetic field), the two renormalization
group (RG) pictures can be studied in a common field theory. This is the generic replica symmetric field
theory elaborated in Refs.\ \cite{rscikk,nucl}. %, see the Appendix where its cubic Lagrangian is displayed.
The vicinity of the (hypothetical) zero temperature
fixed point can be studied in this field theory by assuming a hard (practically infinite) longitudinal
mass, thus projecting the theory into the replicon sector. This was done decades ago by Bray and Roberts
\cite{BrRo}, who found a stable Gaussian fixed point for $d>6$, whereas it was impossible to find
any physically relevant and stable fixed point for $d<6$. This was later interpreted
\cite{BrMo86,Moore_Bray_2011} as a sign that the AT line disappears below six dimensions,
and the droplet scenario takes over. This is, however, a faulty argument, since --- as we have
explained above --- the RG equations (those for instance of Ref.\ \cite{BrRo}) are not specific to the
low temperature behaviour of the AT line. An effort to understand the crossover from the zero-field critical
fixed point to the zero-temperature one was made in Ref.\ \cite{Iveta}, where the whole set of RG equations
was derived in a first order perturbative renormalization. (The Bray-Roberts equations are naturally
included there.) The runaway flows found were discussed in details in \cite{AT2008}, and it was argued in this
reference that the RG scheme used could not be expected to detect a zero-temperature fixed point in epsilon expansion.
But again,
% the failure of finding a zero-temperature fixed point is not specific to the problem of the
%existence of the AT line.
the lack of a fixed point with infinite longitudinal mass {\em in the RG equations valid around the critical
point\/} is not specific to spin glasses, and this property cannot distinguish between the two rival spin
glass theories.
% The problem of the existence or otherwise of the AT line could only be solved by constracting
%RG equations around $T=0$. If these equations possessed a fixed point with a zero value for the magnetic field,
%that would decide the question in favour of the droplet theory, while if the fixed point value of the field
%turned out to be different from zero, that would prove the existence of an AT line.
%\footnote{Since the key for relating a fixed point to either theory (RSB versus droplet)
%is the value of the magnetic field at the fixed point, $h^{*2}$, the Appendix is devoted to presenting the
%generic RG equation (which has not been published before) for $h^2$, and a discussion of different results
%(such as the zero-temperature Gaussian fixed point).} 

In a recent paper \cite{Moore_Bray_2011}, Moore and Bray suggest a proof that RSB disappears when six
dimensions is approached form above. They take the $d\to 6^+$ limit of known first order results,
using RG arguments, for $x_1$ (the breakpoint of the order parameter function) and the AT line, and find
both going to zero. We reproduce their results in a more complete RG scheme in Sec.\ \ref{d>6}, and show
what is the fundamental flaw in their argument. At this point, the reader is advised to jump to
Fig.\ \ref{scaling_variable}(b) in Sec.\ \ref{discussion} where $x_1$ is plotted against dimension
along with the so called scaling variable, which is effectively the relative error of the approximation.
The breakpoint
$x_1$ is monotonically increasing for decreasing dimension as long as the scaling variable is small.
This is the range where the approximation is valid! However at around $d\approx 6.1$, the scaling variable
starts to steeply increase (and actually goes to 1 for $d\to 6$), simultanously $x_1$ suddenly changes
its behaviour, and falls to zero: this is the effect
% that has been found in \cite{Moore_Bray_2011}
(and a similar scenario
for the AT line) that has been found in \cite{Moore_Bray_2011},
but it must be clear that these results fall outside the range of validity of the approximate RG equations.
As a matter of fact, $x_1$ can be calculated directly in $d=6$ (Sec.\ \ref{d=6}), its value is
shown as the horizontal line in Fig.\ \ref{scaling_variable}(b): it is visibly an extrapolation of the
curve from the range where the approximation is good. (In fact, it is an old wisdom of the RG theories
that the upper critical dimension requires special care.) There is only one case where the arguments of
Ref.\ \cite{Moore_Bray_2011} are correct [and interestingly enough, this is admitted there below Eq.\ 
(18) of that reference], namely just at criticality. But that yields only the trivial results for the
$d=6$ system: $x_1$ is zero for $T=T_c$, and the AT line starts at the origin,
i.e.\ at $T=T_c$ and $h^2=0$, and does not say anything about the disappearence of RSB.%
\footnote{Somewhat surprisingly, \cite{Moore_Bray_2011} neglects discussing and even citing
Ref.\ \cite{AT2008}, where the AT line is followed up from mean field to $d\lesssim 6$.
Subsection \ref{AT_below} reconsiders and comfirms the existence of an AT line below six dimensions.}
 
The outline of the paper is as follows: Section \ref{d>6} is devoted to the study of the dimensional
regime $6<d<8$, although the perturbative results of subsection \ref{pert} are extensively used
in later sections too. In Sec.\ \ref{d=6}, the renormalization group ideas
are specifically applied to the $d=6$ case, simply following the lines explained in classical RG textbooks
(see, for instance, \cite{PfeutyToulouse}).
% The renormalization group ideas are specifically applied to the $d=6$ case,
%just along the lines one could have learnt from classical RG textbooks
%(see, for instance, \cite{PfeutyToulouse}), in Sec.\ \ref{d=6}.
The breakpoint $x_1$ and the AT line are calculated
at the upper critical dimension, both displaying logarithmic temperature corrections.
A method for expanding a general (except that replica equivalence is assumed) RSB theory around the RS one is
presented in Sec.\ \ref{RSB_vs_RS}, and applied to the ultrametric case. By this method, quantities of the
RSB theory, like $x_1$, can be expressed in terms of vertices of the RS theory. In the next section,
Sec.\ \ref{d<6},
we return to our original program, and study the case $d<6$: generic RG arguments are presented,
and the calculation of
$x_1$ and the AT line in $\epsilon$-expansion is performed. A new feature emerges below six dimensions,
namely $x_1$ becomes nonzero and universal at criticality. In the last section, Sec.\ \ref{discussion},
special examples, both for $x_1$ and the AT line, are used to conclude that RSB escalates both in the regime
above and below six dimensions. %There is an Appendix where a new {\em anomalous\/} Gaussian fixed point
%of the generic replica symmetric model is presented, and also the renormalization of the magnetic field
%considered.

\section{Formulation of the spin glass problem for $6<d<8$}\label{d>6}

The simplest replicated field theory corresponding to the Ising spin glass
in zero external magnetic field and below $d=8$ has two bare parameters
defining the model: $\tau$ (measuring the distance from criticality and $w$
(the only bare cubic coupling compatible with the symmetrical
--- paramagnetic --- state). Its Lagrangian is
\begin{equation}\label{simple_L}
\mathcal{L}=
\frac{1}{2}\sum_{\mathbf p}
 \bigg(\frac{1}{2} p^2+\bar m\bigg)\sum_{\alpha\beta}
\phi^{\alpha\beta}_{\mathbf p}\phi^{\alpha\beta}_{-\mathbf p}
-\frac{1}{6N^{1/2}}\,\,w\,\sideset{}{'}\sum_{\mathbf {p_1p_2p_3}}
\sum_{\alpha\beta\gamma}\phi^{\alpha\beta}_{\mathbf p_1}
\phi^{\beta\gamma}_{\mathbf p_2}\phi^{\gamma\alpha}_{\mathbf p_3}
\end{equation}
where the bare mass $\bar m=\bar m_c-\tau$, and the critical mass
has been presented in the literature several times in leading order of the
loop expansion:
\[
\bar m_c=\frac{1}{2}(n-2)w^2\frac{1}{N}\sum_{\mathbf p}\frac{1}{p^4}\quad.
\]
In this
$n(n-1)/2$ component field theory the fluctuating fields
are symmetric in the replica indices with zero diagonals:
$\phi^{\alpha\beta}_{\mathbf p}=\phi^{\beta\alpha}_{\mathbf p}$ and
$\phi^{\alpha\alpha}_{\mathbf p}=0$, $\alpha$,$\beta=1,\dots,n$.
[Momentum conservation is indicated by the primed
summation.
The number $N$ of the Ising spins becomes infinite in the thermodynamic
limit, rendering summations to integrals over the continuum of momenta
in the diagrams of the perturbative expansion. A momentum cutoff $\Lambda$
is always understood to block ultraviolet divergences, although it can be
(and will be) absorbed into the definition of different quantities.]
The replica number $n$ goes to zero in the spin glass limit.

\subsection{Perturbative results}\label{pert}

We are now going to recollect several results for the replica symmetric
(RS) spin glass phase --- see
Refs.\ \cite{rscikk,Iveta,nucl,free_energy_fluctuations} --- which are needed
for the following discussion. Due to the severe technical difficulties,
only one-loop calculations have been accomplished ($\epsilon\equiv 6-d$ and $n=0$).
\begin{itemize}
\item RS order parameter $q$, i.e.\ the equation of state:
\begin{equation}\label{q1}
\frac{wq}{\tau}=1-2w^2\,\tau^{|\epsilon|/2}\,\,
\frac{1}{N}\sum_{\mathbf p}^{\frac{\Lambda}{\sqrt{\tau}}}
\frac{p^2-2}{p^4(p^2+2)^2} +\frac{1}{2}w\,\tau^{-2}\,h^2\quad.
\end{equation}
(The last term with the external magnetic field $h$ has been included here
for later reference. At the moment, it is to be considered as zero.)
We can use $wq=\tau$ in the one-loop diagrams, and after rescaling the momentum as
$p\rightarrow p/\sqrt \tau$, two different propagators remain: the replicon
($p^{-2}$) and the longitudinal [$(p^2+2)^{-1}$] ones. To make the formulae for the
one-loop vertices more transparent, it is useful to introduce a common notation
$I_{\dots}$ for the occuring integrals, as is illustrated below:
\[
I_{RRLL}\equiv \frac{1}{N}\sum_{\mathbf p}^{\frac{\Lambda}{\sqrt{\tau}}}
\frac{1}{p^4(p^2+2)^2}=\int^{\frac{\Lambda}{\sqrt{\tau}}}\frac{d^dp}{(2\pi)^d}
\frac{1}{p^4(p^2+2)^2}=K_d\int^{\frac{\Lambda}{\sqrt{\tau}}}
\frac{dp\,p^{-1+d}}{p^4(p^2+2)^2}\quad.
\]
\item The replicon mass:
\begin{equation}\label{GammaR1}
\Gamma_R=2m_1=-2\tau+2wq+4w^2\,\tau^{1+|\epsilon|/2}\,(4I_{RLL}-3I_{RRL})\quad.
\end{equation}
\item The basic cubic vertex of the $\text{Tr}\,\phi^3$ operator:
\begin{equation}\label{w11}
w_1=w+2w^3\,\tau^{|\epsilon|/2}\,(-8I_{RRL}+7I_{RRR}-14I_{RRLL}-8I_{RLLL})\quad.
\end{equation}
\item The quartic vertex of ${\phi^{\alpha\beta}}^4$:
\begin{equation}\label{u21}
u_2=24w^4\,\tau^{-1+|\epsilon|/2}\,I_{RRLL}\quad.
\end{equation}
\end{itemize}
In fact, this last result is new. Details of the somewhat lengthy calculation of the
replicon-type quartic vertices will be published later.

\subsection{Simple two-parameter renormalization group}

An extensive renormalization group (RG) study of the generic RS glassy phase was
published in Ref.\ \cite{Iveta}. When close to the Gaussian fixed point%
\footnote{From now on, we redefine the parameters by suitably absorbing
the geometrical factor $K_d$ and $\Lambda$: $\tau/\Lambda^2\rightarrow
\tau$, $w^2K_d\Lambda^{|\epsilon|}\rightarrow w^2$ and $h^2K_d^{-1/2}
\Lambda^{-4-|\epsilon|/2}\rightarrow h^2$.}%
,
i.e.\ $w\ll 1$ and $\tau\ll 1$, and only infinitesimally breaking the
high-temperature (paramagnetic) symmetry of the system, we have the following
simple two-parameter RG flow-equations:
\begin{equation}\label{RGflow}
\begin{aligned}
\dot{w^2} &=-|\epsilon|w^2-2w^4\,,\\[2pt]
\dot{\tau} &=\left(2-\frac{10}{3}w^2\right)\tau\, .
\end{aligned}
\end{equation}
Physical quantities take simple scaling forms when, instead of $w$ and
$\tau$, they are expressed in terms of the nonlinear scaling fields $\tilde w$ and
$r$ defined by:
\begin{equation}\label{nonlinear}
\begin{aligned}
\dot{\tilde{w}^2\!\!} &=-|\epsilon|{\tilde w}^2\,,\\[2pt]
\dot{r} &=2r\, .
\end{aligned}
\end{equation}
A straightforward calculation provides:
\begin{equation}\label{solution}
\begin{aligned}
%\tilde{w}^2 &=w^2\left(1+2\frac{w^2}{}
w^2&=\tilde{w}^2\left(1-2\frac{\tilde w^2}{|\epsilon|}\right)^{-1}\,,\\[2pt]
\tau &=r\left(1-2\frac{\tilde w^2}{|\epsilon|}\right)^{-5/3}\, .
\end{aligned}
\end{equation}
We are now going to compute the quantities $q$, $\Gamma_R$, $w_1$ and $u_2$
by the RG in terms of $\tilde{w}^2$ and $r$. In this way, we can get more
general results when approaching dimension six from above as compared with
the perturbative computation: now we may have $|\epsilon|\ll w^2\ll 1$,
although the scaling variable $\tilde{w}^2r^{|\epsilon|/2}$ must be small:
\[
\tilde{w}^2r^{|\epsilon|/2}\ll |\epsilon|,\quad\text{even when}\quad
|\epsilon|\ll w^2\,.
\]
\begin{itemize}
\item The renormalization flow equation for $q$ is
\begin{equation}\label{qRG}
\dot{q}=\left(2+\frac{|\epsilon|}{2}+\frac{\eta_L}{2}\right)q
\end{equation}
with $\eta_L=\eta_R=-\frac{2}{3}w^2$ in this approximation.
It can be solved by using Eqs.\ (\ref{nonlinear}) and (\ref{solution}):
\begin{equation}\label{q2}
q=r^{1+\frac{|\epsilon|}{4}}\,\hat q\big(\tilde{w}^2r^{|\epsilon|/2}\big)
\,\left(1-2\frac{\tilde w^2}{|\epsilon|}\right)^{-1/6}\,,
\end{equation}                      
and a comparison with (\ref{q1}) makes it possible --- after some manipulations
--- to get the leading terms of the scaling function:
\begin{equation}\label{q3}
\hat q(x)=\frac{1}{\sqrt{x}}(1+Cx+\dots),\quad \text{with the constant}\quad
C=2^{1+\frac{|\epsilon|}{2}}\,\Gamma\big(1+\frac{|\epsilon|}{2}\big)
\Gamma\big(1-\frac{|\epsilon|}{2}\big)\left(\frac{1}{|\epsilon|}+1\right)\, .
\end{equation}
\item The replicon mass evolves under renormalization as
\begin{equation}\label{GammaR_flow}
\dot{\Gamma}_R=(2-\eta_R)\,\Gamma_R=\left(2+\frac{2}{3}w^2\right)\Gamma_R\, ,
\end{equation}
with the solution
\begin{equation}\label{GammaR2}
\Gamma_R=r\,\hat{\Gamma}_R\big(\tilde{w}^2r^{|\epsilon|/2}\big)
\,\left(1-2\frac{\tilde w^2}{|\epsilon|}\right)^{1/3}\,.
\end{equation}
Substituting $q$ in Eq.\ (\ref{GammaR1}) by $\tau$ from (\ref{q1})  provides:
\begin{equation}\label{GammaR3}
\Gamma_R=-16w^2\,\tau^{1+\frac{|\epsilon|}{2}}\,I_{RRLL}+w\tau^{-1}h^2\,.
\end{equation}
Keeping in mind that (\ref{GammaR3}) is valid for $\tilde w^2\approx w^2
\ll |\epsilon|$ and $h^2$ is zero at the moment, it is straightforward to
derive the scaling function in Eq.\ (\ref{GammaR2}):
\begin{equation}\label{GammaR4}
\hat{\Gamma}_R(x)=C'x+\dots,\quad \text{with}\quad C'=
-2^{2+\frac{|\epsilon|}{2}}\,\Gamma\big(1+\frac{|\epsilon|}{2}\big)
\Gamma\big(1-\frac{|\epsilon|}{2}\big)\,.
\end{equation}
\item As for $w_1$, we have
\begin{equation}\label{w1RG}
\dot{w_1}=\left(-\frac{|\epsilon|}{2}-\frac{3}{2}\eta_R\right)w_1
=\left(-\frac{|\epsilon|}{2}+w^2\right)w_1
\end{equation}
and
\begin{equation}\label{w12}
w_1=r^{-\frac{|\epsilon|}{4}}\,\hat w_1\big(\tilde{w}^2r^{|\epsilon|/2}\big)
\,\left(1-2\frac{\tilde w^2}{|\epsilon|}\right)^{1/2}\,.
\end{equation}
Comparing (\ref{w11}) and (\ref{w12}) yields
\begin{equation}\label{w13}
\hat w_1(x)=\sqrt{x}\,(1+C''x+\dots),\quad \text{with}\quad C''=
2^{\frac{|\epsilon|}{2}}\,\Gamma\big(1+\frac{|\epsilon|}{2}\big)
\Gamma\big(1-\frac{|\epsilon|}{2}\big)\,
\left(\frac{16}{|\epsilon|}-9-|\epsilon|\right)\, .
\end{equation}
\item Finally, from the flow
\begin{equation}\label{u2RG}
\dot{u_2}=\big(-2-|\epsilon|-2\eta_R\big)\,u_2=
\left(-2-|\epsilon|+\frac{4}{3}w^2\right)u_2
\end{equation}
follows the scaling form of the most important quartic vertex:
\begin{equation}\label{u22}
u_2=r^{-1-\frac{|\epsilon|}{2}}\,\hat u_2\big(\tilde{w}^2r^{|\epsilon|/2}\big)
\,\left(1-2\frac{\tilde w^2}{|\epsilon|}\right)^{2/3}\,.
\end{equation}
From (\ref{u21}) and (\ref{u22}) results [see also (\ref{GammaR4})]
\begin{equation}\label{u23}
\hat u_2(x)=-\frac{3}{2}C'x^2+\dots\,.
\end{equation}
\end{itemize}

\subsection{The calculation of $x_1$ and the Almeida--Thouless line}

The leading contribution to the breakpoint of the order parameter function
$q(x)$ is derived in Sec.\ (\ref{RSB_vs_RS}), and has the simple form [see (\ref{x1_basic}) and the
more general considerations in that section about getting $x_1$ on the basis of the generic RS field theory]:
\[
x_1=\frac{u_2}{w_1}\,q\,.
\]
Inserting (\ref{q2}), (\ref{w12}) and (\ref{u22}), the scaling equation of
$x_1$ follows:
\[
x_1=\hat x_1\big(\tilde{w}^2r^{|\epsilon|/2}\big),\quad \text{with}
\quad \hat x_1(\dots)=\frac{\hat u_2(\dots)}{\hat w_1(\dots)}
\,\hat q(\dots)\,.
\]
By the help of Eqs.\ (\ref{q3}), (\ref{w13}), (\ref{u23})
and (\ref{GammaR4}), we can
conclude
\begin{equation}\label{x1_d>6}
x_1=6\times 2^{\frac{|\epsilon|}{2}}\,\Gamma\big(1+\frac{|\epsilon|}{2}\big)
\Gamma\big(1-\frac{|\epsilon|}{2}\big)\,
\,\tilde{w}^2r^{|\epsilon|/2}+\dots\,.
\end{equation}
Inverting (\ref{solution}), $x_1$ can be expressed by the original bare
coupling $w$:
\begin{equation}\label{x1_Moore}
x_1\sim \frac{w^2}{1+2\frac{w^2}{|\epsilon|}}\,r^{|\epsilon|/2}\,.
\end{equation}
This equation agrees with Eq.\ (21) of Ref.\ \cite{Moore_Bray_2011}.%
\footnote{$|r(0)|=|r|$ in that paper is what we call $\tau$ here,
whereas $w(0)=w$ agrees with our notation for the bare cubic coupling.}
The range of applicability of the above equation:
\begin{equation}\label{range}
w^2,r\ll 1,\quad 0<|\epsilon|<2\quad\text{and (most importantly)}\quad
\tilde{w}^2r^{|\epsilon|/2}\ll |\epsilon|\,.
\end{equation}
If we fix the system's bare coupling $w$ and approach six dimensions,
then $\tilde w^2\to |\epsilon|/2$ and $x_1\sim |\epsilon|\,r^{|\epsilon|/2}$.
This behaviour was interpreted by the authors of Ref.\ \cite{Moore_Bray_2011}
as the sign of the end of RSB at six dimensions: a vanishing $x_1$ is
consistent with RS. But, as Eq.\ (\ref{range}) clearly shows, in this limit
$r$ must go to zero,\footnote{That point has been noticed in
Ref.\ \cite{Moore_Bray_2011}, but was completely misinterpreted. We will return to this
problem in Sec.\ \ref{discussion}; see the first row of Eq.\ (\ref{cases}) showing
the impossibility of the limit $|\epsilon|\to 0$ in this approximation.}
i.e.\ the breakpoint disappears at the critical surface in six dimensions
--- a property valid also for $d>6$ (but, as we will see later, not for
$d<6$).\footnote{The multiplicative factor $|\epsilon|$
in $x_1$ has its origin in the termination of the definition of the
nonlinear scaling field $\tilde w$ in $d=6$. This is a feature of the
RS renormalization group, and is not related to the problem of replica symmetry
breaking.} 
In the next section we will show that below the critical surface
$x_1>0$ and has a logarithmic temperature dependence at exactly six dimensions.
%\frac{}{}{\Lambda/\sqrt{\tau}}

We now turn to the problem of the Almeida--Thouless line. The introduction
of a magnetic field $h^2$ involves a new nonlinear scaling field
$\tilde{h^2\!}$ with
\[
\dot{\tilde{h^2\!}}=\left(4+\frac{|\epsilon|}{2}\right)\,\tilde{h^2\!}\,.
\]
Eq.\ (\ref{GammaR2}) remains valid, but the scaling function $\hat \Gamma_R$
has now two arguments: $x=\tilde{w}^2r^{|\epsilon|/2}$ and
$y=\tilde{h^2\!}\,\,r^{-2-|\epsilon|/4}$. Realizing that the replicon mass
starts at one-loop order, the bare parameters in (\ref{GammaR3}) can be replaced
by their corresponding nonlinear scaling fields, making it possible to read
off the scaling function:
\[
\hat \Gamma_R(x,y)=C'x+\sqrt{x}\,y\,;
\]
see also (\ref{GammaR4}). The vanishing replicon mass defines the AT line,
i.e.\ $y=-C'\,\sqrt{x}$ providing
\begin{equation}\label{AT}
\tilde{h^2\!}=-C'\,\tilde w \,r^{2+\frac{|\epsilon|}{2}}\,.
\end{equation}
The connection between $h^2$ and $\tilde{h^2\!}$ may be found from the
flow equation
\begin{equation}\label{h^2_flow}
\dot{h^2}=\left(4+\frac{|\epsilon|}{2}-\frac{\eta_L}{2}\right)\,h^2=
\left(4+\frac{|\epsilon|}{2}+\frac{1}{3}w^2\right)\,h^2\,,
\end{equation}
with the solution [see also (\ref{nonlinear}) and (\ref{solution})]:
\begin{equation}\label{h_solution}
h^2=\tilde{h^2\!}\,\left(1-2\frac{\tilde w^2}{|\epsilon|}\right)^{1/6}\,.
\end{equation}
It is useful to  display the AT line (\ref{AT}) in the original bare parameters
by Eqs.\ (\ref{solution}) and (\ref{h_solution}):
\begin{equation}\label{AT_d>6}
h^2=-C'\, \frac{w}{\left(1+2\frac{w^2}{|\epsilon|}\right)
^{4+\frac{5}{6}|\epsilon|}}
 \,\tau ^{2+\frac{|\epsilon|}{2}}\,.
\end{equation}
This equation is identical with Eq.\ (15) of Ref.\cite{Moore_Bray_2011}, and the
$|\epsilon|^4$ factor, arising when $|\epsilon|\to 0$ while fixing $w$, led those
authors to conclude that the AT line disappears in six dimensions. But,
again, Eq.\ (\ref{range}) and the discussion below it shows that this limit
provides results only on the critical surface ($\tau$ and $r$ zero), and it
informs us only about the trivial fact that the AT line starts at the origin
of the $\tau,h^2$ plain.

\section{At the upper critical dimension: $d=6$}\label{d=6}

As can be seen from the previous section, knowledge about the six dimensional system
cannot be gained from the RG results in the $d\gtrapprox 6$ case.
The fundamental reason for that is the impossibility to linearize the RG flow equations
at exactly an upper critical dimension. Therefore, the scaling field $\tilde w$
is not defined for $d=6$, and we keep $w$ (although $r$ and $\tilde{h^2\!}$
are still meaningful). The RG flow (\ref{RGflow}) is now:
\begin{equation}\label{RGflow_d=6}
\begin{aligned}
\dot{w^2} &=-2w^4\,,\\[2pt]
\dot{\tau} &=\left(2-\frac{10}{3}w^2\right)\tau\, .
\end{aligned}
\end{equation}
The connection between $\tau$ and $r$ becomes [instead of (\ref{solution})]:
\begin{equation}\label{tau_vs_r}
\tau=r\,w^{\frac{10}{3}}\,,
\end{equation}
and the scaling variable with zero scaling dimension is now (instead of
$\tilde{w}^2\,r^{|\epsilon|/2}$):
\[
\frac{w^2}{1-w^2\ln r}\,,
\]
which can be easily checked by Eq.\ (\ref{RGflow_d=6}) and the nonlinear scaling
field property $\dot r=2r$.

\subsection{The calculation of $x_1$}

The renormalization group flow equations for the three relevant physical quantities
$q$, $w_1$ and $u_2$ are as follows:
\begin{gather*}
\dot q= \left(2-\frac{1}{3}w^2\right)\,q\,,\\[4pt]
\dot{w_1}= w^2\,w_1\,,\\[4pt]
\dot{u_2}=\left(-2+\frac{4}{3}w^2\right)\,u_2\,. 
\end{gather*}
They all have the same form, and their solutions are easily found in scaling form.
\begin{itemize}
\item The RS order parameter:
\begin{equation}\label{q_d=6}
\begin{gathered}
q=w^{\frac{1}{3}}\, r\,\hat q\left(\frac{w^2}{1-w^2\ln r}\right)\,,\\[4pt]
\hat q(x)=x\,\left[1+(2+\ln 2)\,x+\frac{5}{3}\,x\ln x+\dots\right]\,.
\end{gathered}
\end{equation}
The scaling function $\hat q(x)$ has been obtained by evaluating (\ref{q1}) in $d=6$
(in zero magnetic field at the moment) and using the connection between
$\tau$ and $r$ in (\ref{tau_vs_r}).
\item The cubic vertex $w_1$ in six dimensions:
\begin{equation}\label{w1_d=6}
\begin{gathered}
w_1=w^{-1}\,\hat w_1\left(\frac{w^2}{1-w^2\ln r}\right)\,,\\[4pt]
\hat w_1=x\,\left[1+\Big(-\frac{39}{2}+8\ln 2-7\ln n\Big)\,x
+\frac{5}{3}\,x\ln x+\dots\right]\,.
\end{gathered}
\end{equation}
Eqs.\ (\ref{w11}) and (\ref{tau_vs_r}) has been used to get the scaling function.
One important remark is appropriate here. The term with the logarithm of the replica
number, $\ln n$, comes from $I_{RRR}$ in (\ref{w11}), and is a prominent example of
the severe infrared divergences caused by the replicon propagator. Similar contributions
enter in higher order vertices, such as $I_{RRRR}$ in the quartic vertex belonging to
the operator $\text{Tr}\,\phi^4$. This is a clear indication --- beside the instability
of the replicon mode --- that the replica symmetric theory is ill-defined in the spin glass
limit. In fact, these infrared divergent terms can be resummed when we build up the RSB
theory on the basis of the RS one, as explained in Sec.\ \ref{RSB_vs_RS}. What is gained in this
resummation, after setting $n$ to zero, is the small mass regime of the RSB solution which
effectively acts as an infrared cutoff. It must be stressed that without this resummation,
the theory is infrared divergent in any arbitrarily high dimension.
\item As for the quartic vertex $u_2$, its scaling form and the leading term of the scaling
function are [see (\ref{u21}) and (\ref{tau_vs_r})]:
\begin{equation}\label{u2_d=6}
\begin{gathered}
u_2=w^{-\frac{4}{3}}\,r^{-1}\,\hat u_2\left(\frac{w^2}{1-w^2\ln r}\right)\,,\\[4pt]
\hat u_2(x)=6x+\dots\,.
\end{gathered}
\end{equation}
\end{itemize}
By Eqs.\ (\ref{q_d=6}), (\ref{w1_d=6}) and (\ref{u2_d=6}) $x_1$ turns out to be a
function of the scaling variable, as it must be:
\[
x_1=\frac{u_2}{w_1}\,q=\hat x_1\left(\frac{w^2}{1-w^2\ln r}\right)
\quad\text{with}\quad \hat x_1(\dots)=\frac{\hat u_2(\dots)}{\hat w_1(\dots)}
\,\hat q(\dots)\,.
\]
The leading order of the scaling function is simply $\hat x_1(x)=6x+\dots$,
providing one of our basic results
\begin{equation}\label{x1_d=6}
x_1=6\,\left(\frac{w^2}{1-w^2\ln r}\right)+\dots\,;\quad w,r\ll 1\quad \text{and}
\quad r=\tau\,w^{-\frac{10}{3}},\quad d=6\,.
\end{equation}
It is clear from the above equation that $x_1$ is zero at criticality
($r=\tau=0$), and for fixed $w$ the approach to zero is logarithmic:
\[
x_1=6\,|\ln r|^{-1}+\dots\,;\quad r,\tau\to 0\quad \text{and}\quad w\quad\text{fixed},\quad d=6\,.
\]

\subsection{Almeida--Thouless line in six dimensions}

The flow equation for the replicon mass is unchanged as compared with the $d>6$ case,
and is given by Eq.\ (\ref{GammaR_flow}). The nonlinear scaling field corresponding to
the external magnetic field satisfies $\dot{\tilde{h^2\!}}=4\,\tilde{h^2\!}$, therefore
the second variable with zero scaling dimension is $\tilde{h^2\!}/r^2$. Straightforward
considerations lead us to
\begin{equation}\label{GammaR_d=6}
 \Gamma_R=w^{-\frac{2}{3}}\, r\,\hat \Gamma_R\left(\frac{w^2}{1-w^2\ln r}\,\,,\,
\frac{\tilde{h^2\!}}{r^2}\right)\,. 
\end{equation}
The evolution of the "bare" magnetic field, i.e.\ 
$\dot{h^2}=\left(4+\frac{1}{3}w^2\right)\,h^2$ [see (\ref{h^2_flow})] and
(\ref{RGflow_d=6}) yield
\begin{equation}\label{h_vs_h}
h^2=\tilde{h^2\!}\,\,w^{-\frac{1}{3}}\,.
\end{equation}
Evaluating Eq.\ (\ref{GammaR3}) at $d=6$, and replacing the bare parameters $\tau$
and $h^2$ by $r$ and $\tilde{h^2\!}$ according to (\ref{tau_vs_r}) and
(\ref{h_vs_h}), respectively, makes it possible to read off the scaling function
in leading order:
\[
\hat \Gamma_R(x,y)=\frac{1}{x}\,(-4x^4+y+\dots)\,.
\]
From its zero, the AT line is obtained as follows:
\begin{equation}\label{AT_d=6}
\tilde{h^2\!}=4\,r^2\,\left(\frac{w^2}{1-w^2\ln r}\right)^4
+\dots\,;\quad w,r\ll 1\quad \text{and}
\quad r=\tau\,w^{-\frac{10}{3}},\quad \tilde{h^2\!}=h^2\,w^{\frac{1}{3}};\quad d=6\,.
\end{equation}
For a given cubic coupling $w$, the magnetic field vs.\ temperature relationship
for the boundary of the RS phase when approaching the critical point becomes:
\[
\tilde{h^2\!}=4\,r^2\,|\ln r|^{-4}+\dots\,;
\quad r,\tau\to 0\quad \text{and}\quad w\quad\text{fixed},\quad d=6\,.
\]

%\frac{}{}   (\ref{}) 
%\bibliography{spinglass}
%\begin{thebibliography}{16}
%\end{thebibliography}
%\end{document}

\section{Formulation of replica symmetry breaking on the basis of the generic
replica symmetric theory}\label{RSB_vs_RS}

The considerations in this section are quite general and, therefore, the paramagnetic
system (i.e.\ an RS system with zero order parameter) 
must be represented --- instead of the simple case of (\ref{simple_L}) which
is sufficient around $d=6$ --- by a model which includes all the invariants
compatible with its higher symmetry \cite{droplet}.
The replicated field theory is now defined by the Lagrangian $\mathcal L$ of the
symmetrical (high-temperature and zero-field) theory:
\begin{multline}\label{L}
\mathcal{L}=
\frac{1}{2}\sum_{\mathbf p}
\bigg(\frac{1}{2} p^2+\bar m_1\bigg)\sum_{\alpha\beta}
\phi^{\alpha\beta}_{\mathbf p}\phi^{\alpha\beta}_{-\mathbf p}
-\frac{1}{6N^{1/2}}\,\sideset{}{'}\sum_{\mathbf {p_1p_2p_3}}
\bar w_1\sum_{\alpha\beta\gamma}\phi^{\alpha\beta}_{\mathbf p_1}
\phi^{\beta\gamma}_{\mathbf p_2}\phi^{\gamma\alpha}_{\mathbf p_3}
-\frac{1}{24N}\,\sideset{}{'}\sum_{\mathbf {p_1p_2p_3p_4}}\\[2pt]
\bigg(\bar u_1\!\!\sum_{\alpha\beta\gamma\delta}\phi^{\alpha\beta}_{\mathbf p_1}
\phi^{\beta\gamma}_{\mathbf p_2}\phi^{\gamma\delta}_{\mathbf p_3}
\phi^{\delta\alpha}_{\mathbf p_4}+\bar u_2\!\sum_{\alpha\beta}
\phi^{\alpha\beta}_{\mathbf p_1}\phi^{\alpha\beta}_{\mathbf p_2}
\phi^{\alpha\beta}_{\mathbf p_3}\phi^{\alpha\beta}_{\mathbf p_4}
+\bar u_3\!\sum_{\alpha\beta\gamma}\phi^{\alpha\gamma}_{\mathbf p_1}
\phi^{\alpha\gamma}_{\mathbf p_2}\phi^{\beta\gamma}_{\mathbf p_3}
\phi^{\beta\gamma}_{\mathbf p_4}+
\bar u_4\!\!\sum_{\alpha\beta\gamma\delta}
\phi^{\alpha\beta}_{\mathbf p_1}\phi^{\alpha\beta}_{\mathbf p_2}
\phi^{\gamma\delta}_{\mathbf p_3}\phi^{\gamma\delta}_{\mathbf p_4}
\bigg)\\[2pt]
-\frac{1}{120N^{3/2}}\,\sideset{}{'}\sum_{\mathbf {p_1p_2p_3p_4p_5}}
\bigg(\bar v_1\!\!\sum_{\alpha\beta\gamma\delta\mu}\phi^{\alpha\beta}_{\mathbf p_1}
\phi^{\beta\gamma}_{\mathbf p_2}\phi^{\gamma\delta}_{\mathbf p_3}
\phi^{\delta\mu}_{\mathbf p_4}\phi^{\mu\alpha}_{\mathbf p_5}+
\bar v_2\!\sum_{\alpha\beta\gamma}
\phi^{\alpha\beta}_{\mathbf p_1}\phi^{\alpha\beta}_{\mathbf p_2}
\phi^{\alpha\beta}_{\mathbf p_3}\phi^{\alpha\gamma}_{\mathbf p_4}
\phi^{\beta\gamma}_{\mathbf p_5}+\\[2pt]
\bar v_3\!\sum_{\alpha\beta\gamma\delta}\phi^{\alpha\beta}_{\mathbf p_1}
\phi^{\beta\gamma}_{\mathbf p_2}\phi^{\gamma\alpha}_{\mathbf p_3}
\phi^{\gamma\delta}_{\mathbf p_4}\phi^{\gamma\delta}_{\mathbf p_5}+
\bar v_4\!\sum_{\alpha\beta\gamma\mu\nu}\phi^{\alpha\beta}_{\mathbf p_1}
\phi^{\beta\gamma}_{\mathbf p_2}\phi^{\gamma\alpha}_{\mathbf p_3}
\phi^{\mu\nu}_{\mathbf p_4}\phi^{\mu\nu}_{\mathbf p_5}
\bigg)+\dots
\end{multline}
where the bare mass $\bar m_1\equiv\bar m=\bar m_c-\tau$,
with $\tau$ measuring the distance from criticality, has been also used in (\ref{simple_L}),
and $\bar w_1\equiv w$ 
(momentum conservation is indicated by the primed summations). The fifth order
invariants with the $\bar v$ bare couplings were also included here. In what follows,
we use the same notation for an {\em exact\/} vertex (e.g.\ $u_2$) and its corresponding
bare coupling ($\bar u_2$), the bar indicating always a bare quantity.

As explained in details in Appendix D of Ref.\ \cite{nucl}, the generic 
Legendre-transformed free energy can be expanded around the RS spin glass state with
order parameter $q$; see (D.5) of this reference:
\begin{multline}\label{F}
\frac{1}{N}\mathcal F(q_{\alpha\beta})=\\[4pt]
\frac{1}{N} \mathcal F(q)+
\frac{1}{2}\left[ m_1\sum_{\alpha\beta}(q_{\alpha\beta}-q)^2+m_2\sum_{\alpha\beta\gamma}
(q_{\alpha\gamma}-q)(q_{\beta\gamma}-q)+m_3\sum_{\alpha\beta\gamma\delta}
(q_{\alpha\beta}-q)(q_{\gamma\delta}-q)\right] \\[4pt]
-\frac{1}{6}\left[w_1\sum_{\alpha\beta\gamma}(q_{\alpha\beta}-q)(q_{\beta\gamma}-q)
(q_{\gamma\alpha}-q)+w_2\sum_{\alpha\beta}(q_{\alpha\beta}-q)^3+
w_3\sum_{\alpha\beta\gamma}(q_{\alpha\beta}-q)^2(q_{\alpha\gamma}-q)+\dots
\right]\\[4pt]
-\frac{1}{24}\left[u_1\sum_{\alpha\beta\gamma\delta}(q_{\alpha\beta}-q)(q_{\beta\gamma}-q)
(q_{\gamma\delta}-q)(q_{\delta\alpha}-q)
+u_2\sum_{\alpha\beta}(q_{\alpha\beta}-q)^4
%+u_3\sum_{\alpha\beta\gamma}(q_{\alpha\gamma}-q)^2(q_{\beta\gamma}-q)^2
+\dots\right]\\[4pt]
-\frac{1}{120}\left[v_1\sum_{\alpha\beta\gamma\delta\mu}(q_{\alpha\beta}-q)(q_{\beta\gamma}-q)
(q_{\gamma\delta}-q)(q_{\delta\mu}-q)(q_{\mu\alpha}-q)
%+v_2\sum_{\alpha\beta\gamma}(q_{\alpha\beta}-q)^3(q_{\alpha\gamma}-q)
%(q_{\beta\gamma}-q)
+\dots\right]+\dots\quad.
\end{multline}
In zero external field $\mathcal F(q_{\alpha\beta})$ has the same symmetry as
$\mathcal L$ of Eq.\ (\ref{L})
--- which is higher than that of a generic RS system ---,
even when $T<T_c$, and using this symmetry, a set of
equations can be found between the exact vertices of the generic RS theory (see
Refs.\ \cite{droplet,nucl}). The most effective way to get the required vertex
relationships is demanding that invariants incompatible with the symmetrical theory,
e.g.\ $\sum_{\alpha\beta}q_{\alpha\beta}^3$, must finally disappear from
(\ref{F}). In this manner, all the vertices of the lower symmetry: $m_2$, $m_3$;
$w_2$,\dots $w_8$; $u_5$,\dots $u_{23}$; \dots etc., (see Appendix A of
Ref.\ \cite{nucl} for
the classification of cubic and quartic vertices) and, as a bonus, $m_1$ can be 
expressed in terms of $w_1$, $u_1$, $u_2$, $u_3$, $u_4$, $v_1$, $v_2$, $v_3$, $v_4$,
and higher order symmetrical vertices.%
\footnote{A vertex is called symmetrical if it is nonzero in the zero order parameter
RS system.}
We than have
\begin{align*}
m_1&=\frac{1}{2}nw_1\,q+\frac{1}{6}(n^2u_1-2u_2)\,q^2+\frac{1}{24}
n(n^2v_1-2v_2)\,q^3+\dots,\\[3pt]
m_2&=-w_1\,q-\frac{1}{3}(nu_1+u_3)\,q^2+\frac{1}{60}[5n(3n^2-5n+1)v_1+2v_2-4nv_3]
\,q^3+\dots,\\[3pt]
m_3&=-\frac{1}{6}(u_1+2u_4)\,q^2-\frac{1}{60}[5(5n-4)v_1+2v_3+6nv_4]\,q^3+\dots,%\\[3pt]
\end{align*}
and furthermore
\begin{align*}
w_2&=u_2\,q+\frac{1}{20}nv_2\,q^2+\dots,\\[3pt]
w_3&=u_3\,q+\frac{1}{10}(3v_2+nv_3)\,q^2+\dots,\\[3pt]
w_4&=u_4\,q+\frac{1}{20}(v_3+3nv_4)\,q^2+\dots,\\[3pt]
w_5&=u_1\,q+\frac{1}{20}(5nv_1+4v_3)\,q^2+\dots,\\[3pt]
w_6&=\frac{1}{10}v_3\,q^2+\dots,\\[3pt]
w_7&=\frac{1}{40}(10v_1+12v_4)\,q^2+\dots,\\[3pt]
w_8&=O(q^3).
\end{align*}
Of the quartic vertices, only those are listed below which are required 
up to the order of the present calculation:
\begin{align*}
u_5&=\frac{3}{5}v_2\,q+\dots,&  u_6&=\frac{2}{5}v_3\,q+\dots,&  u_7&=\frac{2}{5}v_4\,q+\dots,\\[3pt]
u_8&=\frac{2}{5}v_2\,q+\dots,&  u_{10}&=\frac{1}{5}v_3\,q+\dots,&
u_{11}&=\frac{2}{5}v_3\,q+\dots,\\[3pt]
u_{14}&=\frac{3}{5}v_4\,q+\dots,&  u_{16}&=v_1\,q+\dots.&&
\end{align*} 
By exploiting these expressions, the free energy functional in Eq.\ (\ref{F}) can now be written
(omitting an additive term depending only on $q$):
\begin{multline*}
\frac{1}{N}\mathcal F(q_{\alpha\beta})=\frac{1}{4}Mq\,\sum_{\alpha\beta}q_{\alpha\beta}^2-
\frac{1}{6}W\,\sum_{\alpha\beta\gamma}q_{\alpha\beta}q_{\beta\gamma}q_{\gamma\alpha}
-\frac{1}{24}\Big[(u_1+v_1q+\dots)\,\sum_{\alpha\beta\gamma\delta}
q_{\alpha\beta}q_{\beta\gamma}q_{\gamma\delta}q_{\delta\alpha}\\
+\big(u_2+\frac{2}{5}v_2q+\dots\big)\,\sum_{\alpha\beta}q_{\alpha\beta}^4+
\big(u_3+\frac{3}{5}v_3q+\dots\big)\,\sum_{\alpha\beta\gamma}q_{\alpha\gamma}^2q_{\beta\gamma}^2
+\big(u_4+\frac{3}{5}v_4q+\dots\big)\,\Big(\sum_{\alpha\beta}q_{\alpha\beta}^2\Big)^2\Big]\\
-\frac{1}{120}\Big[(v_1+\dots)\,\sum_{\alpha\beta\gamma\delta\mu}q_{\alpha\beta}q_{\beta\gamma}
q_{\gamma\delta}q_{\delta\mu}q_{\mu\alpha}+(v_2+\dots)\,\sum_{\alpha\beta\gamma}q_{\alpha\beta}^3
q_{\alpha\gamma}q_{\beta\gamma}+(v_3+\dots)\,\sum_{\alpha\beta\gamma\delta}q_{\alpha\beta}
q_{\beta\gamma}q_{\gamma\alpha}q_{\gamma\delta}^2\\
+(v_4+\dots)\,\Big(\sum_{\alpha\beta\gamma}
q_{\alpha\beta}q_{\beta\gamma}q_{\gamma\alpha}\Big)\Big(\sum_{\alpha\beta}q_{\alpha\beta}^2\Big)
\Big]-\dots\quad,
\end{multline*}
with the following notations for $M$ and $W$:
\begin{align*}
M&\equiv (n-2)w_1\\[3pt]
&\!\!\!+\frac{1}{3}[(n^2-3)u_1+u_2+(n-1)\tilde u_3]\,q+
\frac{1}{60}[5(n^3-4)v_1-2(2n-8)v_2+2(n-1)(n+4)\tilde v_3]\,q^2+\dots,\\[3pt]
W&\equiv w_1+u_1\,q+\frac{1}{20}[10v_1-3v_2-(n-1)\tilde v_3]\,q^2+\dots\quad.
\end{align*}
The tilded vertices $\tilde u_3\equiv u_3+nu_4$ and $\tilde v_3\equiv v_3+nv_4$ were introduced
here; in fact, only these combinations will enter the equation of state.

Stationarity of the free energy functional provides the equation of state:
\begin{multline}\label{basic}
0=Mq\,q_{\alpha\beta}-W\,(q^2)_{\alpha\beta}-\frac{1}{3}\big[(u_1+v_1q+\dots)\,(q^3)_{\alpha\beta}+
(u_2+\frac{2}{5}v_2q+\dots)\,q_{\alpha\beta}^3\\[4pt]
+(\tilde u_3+\frac{3}{5}\tilde v_3q+\dots)\,
(q^2)_{\alpha\alpha}\,q_{\alpha\beta}\big]%\\
-\frac{1}{60}\Big\{5(v_1+\dots)\,(q^4)_{\alpha\beta}+(v_2+\dots)\,\big[3(q^2)_{\alpha\beta}\,q_{\alpha\beta}^2+
\sum_{\gamma}(q_{\alpha\gamma}^3q_{\beta\gamma}+q_{\beta\gamma}^3q_{\alpha\gamma})\big]\\[2pt]+
(\tilde v_3+\dots)\,\big[2(q^3)_{\alpha\alpha}\,q_{\alpha\beta}+3(q^2)_{\alpha\alpha}\,
(q^2)_{\alpha\beta}\big]\Big\}-\dots\quad.
\end{multline}
Only replica equivalence was used in the derivation of this equation --- $(q^2)_{\alpha\alpha}$, for instance,
is independent of the replica number ---, otherwise it is quite general: it provides an RSB solution in terms
of the RS order parameter $q$ (which measures the distance from criticality now), and of the exact RS vertices.
It can equally be used in any regime where some kind of perturbation theory is valid.

We now turn to the case of infinite step, ultrametrically organized RSB. The small parameter making possible a
perturbative treatment is $x_1$, the breakpoint of the order parameter function: it is proportional to $q$
%($q^{d/2-3}$)
in the SK model and for the field theory above 8 dimensions, to $q^{d/2-3}$ between 6 and 8 dimensions,
% (between 6 and 8 dimensions),
whereas it is of order
$\epsilon$ below 6 dimensions. $q(x)$, the order parameter function, has the form:\footnote{We hope that
the {\em ratio}\/ $r$ of $x$ to $x_1$ introduced here cannot be confused with the temperature-like
scaling field of previous sections.}
\begin{equation}\label{q(x)}
q(x)=q_1\,[r+x_1^2\,\delta\bar q(r)],\qquad \text{with}\qquad r\equiv x/x_1\qquad\text{and}\qquad
\delta\bar q(1)=0.
\end{equation}

The contributions of the various vertices to Eq.\ (\ref{basic}) are listed below.
The definition of the
bilinear expression $\{\dots;\dots\}$ used extensively in that list is as follows:
\[
\{f(r);g(r)\}\equiv f(r)g(1)+f(1)g(r)-f(r)\int_r^1du g(u) -g(r)\int_r^1du f(u)-rf(r)g(r)
-\int_0^rdu f(u)g(u).
\]
\begin{itemize} 
\item $w_1$:
\[
2(q_1-q)\,r-x_1q_1\,(r-\frac{1}{3}r^3)+2x_1^2(q_1-q)\,\delta\bar q(r)-2x_1^3q_1\,\{r;\delta\bar q(r)\}+O(x_1^4),
\]
\item $u_1$:
\[
-(q_1-q)^2\,r+x_1(q_1-q)q_1\,(r-\frac{1}{3}r^3)-\frac{1}{3}x_1^2q_1^2\,(\frac{3}{4}r-\frac{1}{2}r^3+\frac{3}{20}r^5)
+O(x_1^4),
\]
\item $u_2$:
\[
\frac{1}{3}q^2\,r-\frac{1}{3}q_1^2\,r^3+\frac{1}{3}x_1^2q^2\,\delta\bar q(r)-x_1^2q_1^2\,r^2\delta\bar q(r)
+O(x_1^4),
\]
\item $\tilde u_3$:
\[
\frac{1}{3}(q_1^2-q^2)\,r-\frac{2}{9}x_1q_1^2\,r+\frac{1}{3}x_1^2(q_1^2-q^2)\,\delta\bar q(r)
-\frac{2}{3}x_1^3q_1^2\,r\{r;\delta\bar q(r)\}_{r=1}-\frac{2}{9}x_1^3q_1^2\,\delta\bar q(r)+O(x_1^4),
\]
\item $v_1$:
\begin{multline*}
\frac{1}{3}(q_1-q)^3\,r-\frac{1}{2}x_1(q_1-q)^2q_1\,(r-\frac{1}{3}r^3)+\frac{1}{3}x_1^2(q_1-q)
q_1^2\,(\frac{3}{4}r-\frac{1}{2}r^3+\frac{3}{20}r^5)\\[5pt]
-\frac{1}{12}x_1^3q_1^3\,(\frac{1}{2}r-\frac{1}{2}r^3
+\frac{3}{10}r^5-\frac{1}{14}r^7)+O(x_1^4),
\end{multline*}
\item $v_2$:
\begin{multline*}
-\frac{1}{5}(q_1-q)q^2\,r+\frac{2}{15}(q_1-q)q_1^2\,r^3
-\frac{1}{30}x_1q_1^3\,(\frac{3}{4}r+\frac{1}{2}r^3-\frac{9}{20}r^5)
+\frac{3}{20}x_1q^2q_1\,(r-\frac{1}{3}r^3)\\[5pt]
-\frac{1}{20}x_1q_1^3\,r^2(r-\frac{1}{3}r^3)+\frac{1}{10}(q_1-q)^2q\,r+\frac{1}{30}(q_1-q)^3\,r
-\frac{1}{5}x_1^2(q_1-q)q_1^2\,(1-2r^2)\delta\bar q(r)\\[5pt]
-\frac{1}{10}x_1^3q_1^3\,r(r-\frac{1}{3}r^3)\delta\bar q(r)+\frac{3}{10}x_1^3q_1^3\,\{r;\delta\bar q(r)\}
-\frac{1}{10}x_1^3q_1^3\,r^2\{r;\delta\bar q(r)\}-\frac{1}{30}x_1^3q_1^3\,\{r^3;\delta\bar q(r)\}\\[5pt]
-\frac{1}{10}x_1^3q_1^3\,\{r;r^2\delta\bar q(r)\}+O(x_1^4),
\end{multline*}
\item $\tilde v_3$:
\begin{multline*}
\Big[-\frac{3}{10}(q_1-q)^2q_1+\frac{2}{15}x_1(q_1-q)q_1^2-\frac{1}{75}x_1^2q_1^3
+\frac{2}{15}(q_1-q)^3\Big]\,r\\[5pt]
+\frac{1}{20}x_1q_1\Big[2(q_1-q)q_1-\frac{2}{3}x_1q_1^2-(q_1-q)^2\Big]\,(r-\frac{1}{3}r^3)+O(x_1^4).
\end{multline*}
\end{itemize}

%$x_1$ is found from Eq.\ (\ref{basic}), and by the above expressions, as the zero, $f(x_1)=0$, of the following
%function:
Inserting the above expressions into Eq.\ (\ref{basic}) and demanding that the coefficients of $r$ and
$r^3$ disappear, $x_1$ can be read off with some effort. It is best to give $x_1$ as the zero, $f(x_1)=0$,
of the following function:
\begin{multline*}
f(x)\equiv\\
 \Big[-\big(\frac{u_2}{w_1}q\big)+\frac{1}{2}\big(\frac{y_2}{w_1}q^3\big)+\dots\Big]
+\Big[1-\frac{13}{60}\big(\frac{v_2}{w_1}q^2\big)+\dots\Big]\,x+\Big[-\frac{1}{3}+\frac{1}{6}
\big(\frac{u_1}{w_1}q\big)+\dots\Big]\,x^2+\Big[-\frac{1}{9}+\dots\Big]\,x^3+\dots\,.%O(x^4).
\end{multline*}
The leading contribution is the well-known formula
\begin{equation}\label{x1_basic}
x_1=\frac{u_2}{w_1}q
\end{equation}
which is used extensively throughout this paper.
As a byproduct, the shift of $q_1$ from the RS order parameter is given by
\begin{equation}\label{q_1}
q_1-q=\frac{1}{3}\,x_1\,q\big(1+\frac{2}{3}x_1+\dots\big)\quad.
\end{equation}
[To preserve consistency, a sixth order contribution
$-\frac{1}{6!}\,y_2\,\sum_{\alpha\beta}(q_{\alpha\beta}-q)^6$ should have been included in the free energy
expansion (\ref{F}), as it enters the constant of $f(x)$ at the third order, i.e.\ at the highest order
studied here.]

\section{Below six dimensions}\label{d<6}

\subsection{The renormalization group: fixed point and nonlinear scaling fields}
In $d=6-\epsilon$ the Gaussian fixed point becomes unstable, and the zero
field spin glass transition is governed by the non-trivial one. Here we collect and present the
available results for this fixed point (in the results for the fixed point below, a
generic $n$ is kept, although $n=0$ is taken in the further parts of the section)
:
\[
\bar {w}_1^{*2}\equiv w^{*2}=\frac{1}{2-n}\,\epsilon\,,\quad
\bar{m}_1^*\equiv \bar{m}^*=-\frac{2-n}{4}\,{w^*}^2\,;\quad \text{see 
Refs.\ \cite{HaLuCh76} and \cite{Iveta}}\,.
\]
Although they will not be used in this paper, the fixed point values of the quartic couplings
(which --- according to our knowledge --- have not been published before) are also listed here:
\[
\bar{u}_1^*=\frac{3}{2}n\,w^{*4},\quad\bar{u}_2^*=12\,w^{*4},
\quad\bar{u}_3^*=-24\,w^{*4}, \quad\bar{u}_4^*=\frac{9}{2}\,w^{*4}\,.
\]
The renormalization flow equations for the bare couplings of the generic
RS theory were displayed in Ref.\ \cite{Iveta}. Using these equations, a new set of parameters
$g_i$
--- the so called nonlinear scaling fields introduced by Wegner \cite{Wegner} --- can
be defined with the following properties:%
\footnote{The summary presented in this paragraph about the use of nonlinear scaling fields
is quite general, not limited to the nontrivial fixed point of the RS replica field theory.}
\begin{itemize}
\item $g_i\equiv 0$ at the fixed point for all $i$.
\item An infinitesimally small $g_i$, with all the others kept zero, gives an eigenvector
belonging to the eigenvalue $\lambda_i$
of the linearized renormalization group equations around the fixed point.
\item They satisfy {\em exactly\/} the equations $\dot g_i=\lambda_i\,g_i$.
\end{itemize}
The RG flow of an observable $y$ --- the order parameter or an irreducible vertex,
for instance --- can be written in terms of the $g_i$'s as follows:
\begin{equation}\label{yRG}
\dot y=\Big(k+\sum_i k_i\,g_i+\sum_{ij} k_{ij}\, g_ig_j+\dots\Big)\,\,y\quad .
\end{equation}
The solution of this equation, i.e.\ $y$ in terms of the scaling fields is
easily found:
\begin{equation}\label{y}
y(g_1,g_2,\dots)=g_1^{k/\lambda_1}\,
\hat y\Big(g_2\,g_1^{-\lambda_2/\lambda_1},\dots,g_i\,g_1^{-\lambda_i/\lambda_1},\dots\Big)
\times \exp\Big(\sum_i \frac{k_i}{\lambda_i}\,g_i
+\sum_{ij} \frac{k_{ij}}{\lambda_i+\lambda_j}\, g_ig_j+\dots\Big)\,,
\end {equation}
the scaling function $\hat y(\dots)$ can be determined by perturbative methods.

In our two-parameter system defined by $\tau$ and $w$ the two nonzero scaling fields%
\footnote{In references \cite{droplet,AT2008} an alternative scheme was used with a second
relevant scaling field entering after appropriately redefining the field theory
for getting rid of "tadpole" diagrams. The irreducible vertices are the same in the two
schemes.}
$r\equiv g_1$ and $\tilde g\equiv g_2$
(the notations are chosen to keep connection with previous sections)
can be found by starting with the RG equations (\ref{RGflow})%
\footnote{\label{note}But be careful to replace $-|\epsilon|$ with $\epsilon$.}
and taking the temperature-like relevant eigenvalue $\lambda_r$ and
the irrelevant one, $\lambda_{\tilde g}$, from Ref.\ \cite{Iveta} as
\begin{equation}\label{lambda}
\lambda_r\equiv \frac{1}{\nu}=2-\frac{5}{3}\,\epsilon+\dots\,,\qquad\qquad
\lambda_{\tilde g}=-\epsilon+\dots\,.
\end{equation}
The bare parameters are then straightforwardly expressed by the scaling fields as
\begin{equation}\label{r_g}
\begin{aligned}
%\tau&=r\,(1-2\frac{\tilde g}{\epsilon})^{-5/3}\\[3pt]
w^2&=w^{*2}+\frac{\tilde g}{1-2\frac{\tilde g}{\epsilon}}=
\frac{\epsilon/2}{1-2\frac{\tilde g}{\epsilon}}\\[5pt]%\,\quad.
\tau&=r\,\Big(1-2\frac{\tilde g}{\epsilon}\Big)^{-5/3}\,.
\end{aligned}
\end{equation}

\subsection{$x_1$ below the upper critical dimension}

For the calculation of $x_1$ to first order in $\epsilon$, the RG study of $q$, $w_1$
and $u_2$ is required. The truncated (one-loop) renormalization group equations
(\ref{qRG}), (\ref{w1RG}) and (\ref{u2RG}) --- see also footnote \ref{note} ---
can be used whenever $w^2\ll 1$ and $\tau \ll 1$. We can solve these truncated equations
in a similar way as
(\ref{y}) was derived from the generic equation (\ref{yRG}).
The scaling exponents and the relations between bare and scaling parameters
are taken from Eqs.\ (\ref{lambda}) and (\ref{r_g}), respectively. The scaling functions,
which are always denoted by the "hat" symbol, cannot be determined by the RG equations
alone, but the perturbative results of Eqs.\ (\ref{q1}), (\ref{w11})
and (\ref{u21}) make it possible
to get them to first order in $\epsilon$. [The bare values must be replaced by the scaling
fields using (\ref{r_g}), and take into account again footnote \ref{note}.] In the
following, the results for $q$, $w_1$ and $u_2$ are listed in itemized form. The
$k$ and $k_2\equiv k_{\tilde g}$ quantities defined in (\ref{yRG}) are also presented for
completeness.
% the generic equation (\ref{yRG}) in the last subsection provided
%(\ref{y}): the scaling exponents and the relations between bare and scaling parameters
%are taken from Eqs.\ (\ref{lambda}) and (\ref{r_g}), respectively. 
\begin{itemize}
\item
\begin{gather}
q=r^{1+\frac{\epsilon}{2}}\,\,\hat q\big(\tilde g\,r^{\frac{\epsilon}{2}}\big)
\times\Big(1-2\,\frac{\tilde g}{\epsilon}\Big)^{-1/6}\,,\label{q_scaling}\\[4pt]
\intertext{with the scaling function}
w^*\hat q(x)=\Big[1+\big(\frac{1}{2}\ln 2+1\big)\,\epsilon+\dots\Big]+
2\,\Big[1+\big(\ln 2+\frac{17}{6}\big)\,\epsilon+\dots\Big]\,\,
\Big(\frac{x}{\epsilon}\Big)+O\Big[\Big(\frac{x}{\epsilon}\Big)^2\Big]\,;
\label{q_hat}\\[10pt]
k=2-\frac{\epsilon}{2}+\frac{1}{2}\,\eta^*_L=2-\frac{2}{3}\,\epsilon+O(\epsilon^2)
\qquad\text{and}\qquad k_{\tilde g}=-\frac{1}{3}+O(\epsilon)\,.
\label{q_k}%\notag
\end{gather}
\item
\begin{gather}
w_1=r^{\frac{\epsilon}{2}}\,\,\hat w_1\big(\tilde g\,r^{\frac{\epsilon}{2}}\big)
\times\Big(1-2\,\frac{\tilde g}{\epsilon}\Big)^{1/2}\,,
\label{w1_scaling}\\[4pt]
\intertext{with the scaling function}
\frac{\hat w_1(x)}{w^*}=\Big[1+\big(4\ln 2-\frac{39}{4}-\frac{7}{2}\ln n\big)\,\epsilon+\dots\Big]+
2\,\Big[1+\big(8\ln 2-\frac{56}{3}-7\ln n\big)\,\epsilon+\dots\Big]\,\,
\Big(\frac{x}{\epsilon}\Big)\notag
\\[10pt]+O\Big[\Big(\frac{x}{\epsilon}\Big)^2\Big]\,;\qquad\qquad\qquad\qquad
\qquad\qquad\qquad\qquad\qquad\qquad\qquad\qquad\,\,
\label{w1_hat}\\[10pt]
k=\frac{\epsilon}{2}-\frac{3}{2}\,\eta^*_R=\epsilon+O(\epsilon^2)
\qquad\text{and}\qquad k_{\tilde g}=1+O(\epsilon)\,.
\notag
\end{gather}
\item
\begin{gather}
u_2=r^{-1}\,\,\hat u_2\big(\tilde g\,r^{\frac{\epsilon}{2}}\big)
\times\Big(1-2\,\frac{\tilde g}{\epsilon}\Big)^{2/3}\,,
\label{u2_scaling}\\[4pt]
\intertext{with the scaling function}
\frac{\hat u_2(x)}{w^{*4}}=6\,\Big[1+O(\epsilon)\Big]+12\,\Big[1+O(\epsilon)\Big]
\,\,\Big(\frac{x}{\epsilon}\Big)+O\Big[\Big(\frac{x}{\epsilon}\Big)^2\Big]\,;
\label{u2_hat}\\[10pt]
k=-2+\epsilon-2\,\eta^*_R=-2+\frac{5}{3}\,\epsilon+O(\epsilon^2)
\qquad\text{and}\qquad k_{\tilde g}=\frac{4}{3}+O(\epsilon)\,.
\notag
\end{gather}
\end{itemize}
The following remarks are appropriate here: Firstly, according to Eqs.\ (\ref{y}),
(\ref{lambda}) and (\ref{q_k}) the temperature
exponent for the order parameter $q$ is {\em exactly\/} $k/\lambda_r=(2-\epsilon/2+\eta^*/2)\,\nu
\equiv \beta=1+\epsilon/2+O(\epsilon^2)$; see (\ref{q_scaling}). The temperature exponents
in (\ref{q_scaling}), (\ref{w1_scaling}) and (\ref{u2_scaling}) are correct up to $\epsilon$ order.
Secondly, the discussion below Eq.\ (\ref{w1_d=6}) concerning the fully replicon, infrared
divergent contribution to $w_1$ is equally valid for the $\ln n$ terms in (\ref{w1_hat}).
Thirdly, the $O(\epsilon)$ corrections in the scaling functions $\hat q$ and $\hat w_1$ are
unnecessary for the leading order calculation of $x_1$; they are displayed here to demonstrate
the general form of the $\epsilon$ expansion. The corresponding corrections for $\hat u_2$ are
not even available, see (\ref{u2_hat}), as they would require a two-loop level calculation.

The leading contribution in the $\epsilon$ expansion for $x_1$ follows from substituting
$q$, $w_1$ and $u_2$ from Eqs.\ (\ref{q_scaling})-(\ref{q_hat}), (\ref{w1_scaling})-(\ref{w1_hat})
and (\ref{u2_scaling})-(\ref{u2_hat}), respectively, into the basic formula in Eq.\ (\ref{x1_basic}).
A remarkably simple formula reflecting the invariance of $x_1$ under renormalization can be
concluded:
\begin{equation}\label{x1_d<6}
\begin{gathered}
x_1=6\,w^{*2}\,\hat x_1\big(\tilde g\,r^{\frac{\epsilon}{2}}\big)
=3\,\epsilon \,\hat x_1\big(\tilde g\,r^{\frac{\epsilon}{2}}\big)\,,\quad
\text{ with the scaling function}\\[6pt]
\hat x_1(x)=\Big[1+O(\epsilon)\Big]+2\,\Big[1+O(\epsilon)\Big]
\,\,\Big(\frac{x}{\epsilon}\Big)+O\Big[\Big(\frac{x}{\epsilon}\Big)^2\Big]\,.
\end{gathered}
\end{equation}

\subsection{Almeida-Thouless line for $d<6$}\label{AT_below}

The external magnetic field $h^2$ evolves under renormalization according to Eq.\ (\ref{h^2_flow}),
with $|\epsilon|$ replaced by $-\epsilon$. The corresponding nonlinear scaling field $g_3\equiv\tilde{h^2}$
has now the relevant eigenvalue
\[
\lambda_{\tilde{h^2}}=4-\frac{\epsilon}{2}-\frac{\eta^*}{2}\equiv \frac{\delta\,\beta}{\nu}\,.
\]
The flow equation for the replicon mass --- Eq.\ (\ref{GammaR_flow}) --- does not contain explicitly
the magnetic field, therefore it enters the solution only through the invariant
$\tilde{h^2}\,r^{-\delta\,\beta}$; see Eqs.\ (\ref{yRG}), (\ref{y}) and (\ref{lambda}).
According to the generic scheme (\ref{y}), we have
\[
\Gamma_R=r^{(2-\eta^*)\,\nu}\,\,\hat\Gamma_R\big(\tilde g\,r^{-\lambda_{\tilde g}\,\nu},
\tilde{h^2}\,r^{-\delta\,\beta},\dots\big)\times\exp \Big(\frac{2}{3}\,\frac{\tilde g}{\lambda_{\tilde g}}
+\dots\Big)\,.
\]
The exponential part can again be calculated in the truncated, one-loop approximation, in the usual
way, providing (note that $\lambda_{\tilde g}=-\epsilon+\dots$)
\[
\Big (1-2\,\frac{\tilde g}{\epsilon}\Big)
^\frac{1}{3}\,,
\]
whereas a comparison with the perturbative result (\ref{GammaR3}) --- after substituting the bare parameters
by their corresponding nonlinear scaling fields [see Eq.\ (\ref{r_g}) and also
\begin{equation}\label{h}
h^2=\tilde{h^2}\,\Big (1-2\,\frac{\tilde g}{\epsilon}\Big)
^{1/6}
\end{equation}
which follows from (\ref{h^2_flow})] --- gives the scaling function:
\begin{equation}\label{hat_GammaR}
\hat\Gamma_R(x,y)=w^{*2}\,\Big\{[-4+O(\epsilon)]+[-24+O(\epsilon)]\,\Big(\frac{x}{\epsilon}\Big)
+[1+O(\epsilon)]\,\Big(\frac{y}{w^*}\Big)+[-2+O(\epsilon)]\,\Big(\frac{x}{\epsilon}\Big)
\,\Big(\frac{y}{w^*}\Big)+\dots
\Big\}\,.
\end{equation}
%It may be useful to show the range of validity of the above expansion for the scaling function:
%\[
%\frac{x}{\epsilon}=\frac{2}{\epsilon}\,\tilde g\,r^{\epsilon/2}\ll 1\,\quad
%
%\]
The zero of the scaling function gives the AT-line:
\begin{equation}\label{AT_d<6}
\tilde{h^2}=4\,w^*r^{\delta\,\beta}=4\,w^*r^{2+\dots}\,,\qquad \frac{\tilde g}{\epsilon}\,
r^{\epsilon/2}\ll 1\,\quad\text{and}\,\quad 0<\epsilon\ll 1\,.
\end{equation}
For the fixed point system, $w=w^*$ implies $\tilde g=0$ and $\tilde{h^2}=h^2$, $r=\tau$.
The result in (\ref{AT_d<6}) is then identical with Eq.\ (18) of Ref.\ \cite{AT2008}.

%\frac{}{}   (\ref{})
\section{Discussion of the results and conclusions}

\label{discussion}
\begin{figure}\caption{The scaling variable (left vertical axis) 
measures the goodness of the approximation. (a): $d<6$ and (b): $d>6$.
The dependence of $x_1$ is also shown in both regimes, together with its
$d=6$ value (horizontal lines). $w^2=0.005$ and $\tau=0.0001$ are fixed in this figure.
The approximation breaks down when approaching $d=6$ from both sides.}
\label{scaling_variable}
\vspace{20pt}
%\fbox{%
\includegraphics[bb=20 118 575 673,scale=0.3]{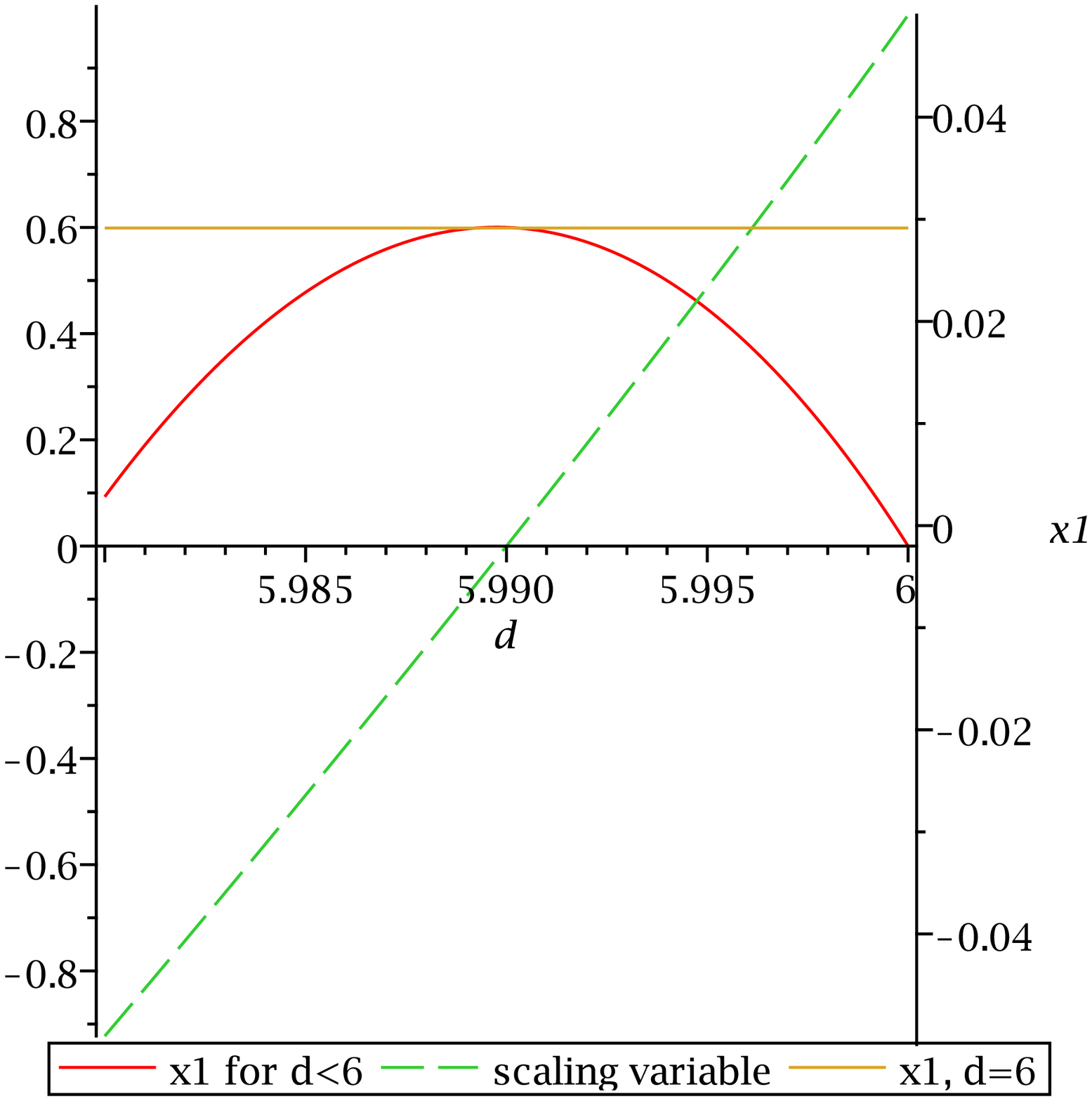}%
%}
\qquad\qquad
%\fbox{%
\includegraphics[bb=20 118 575 673,scale=0.3]{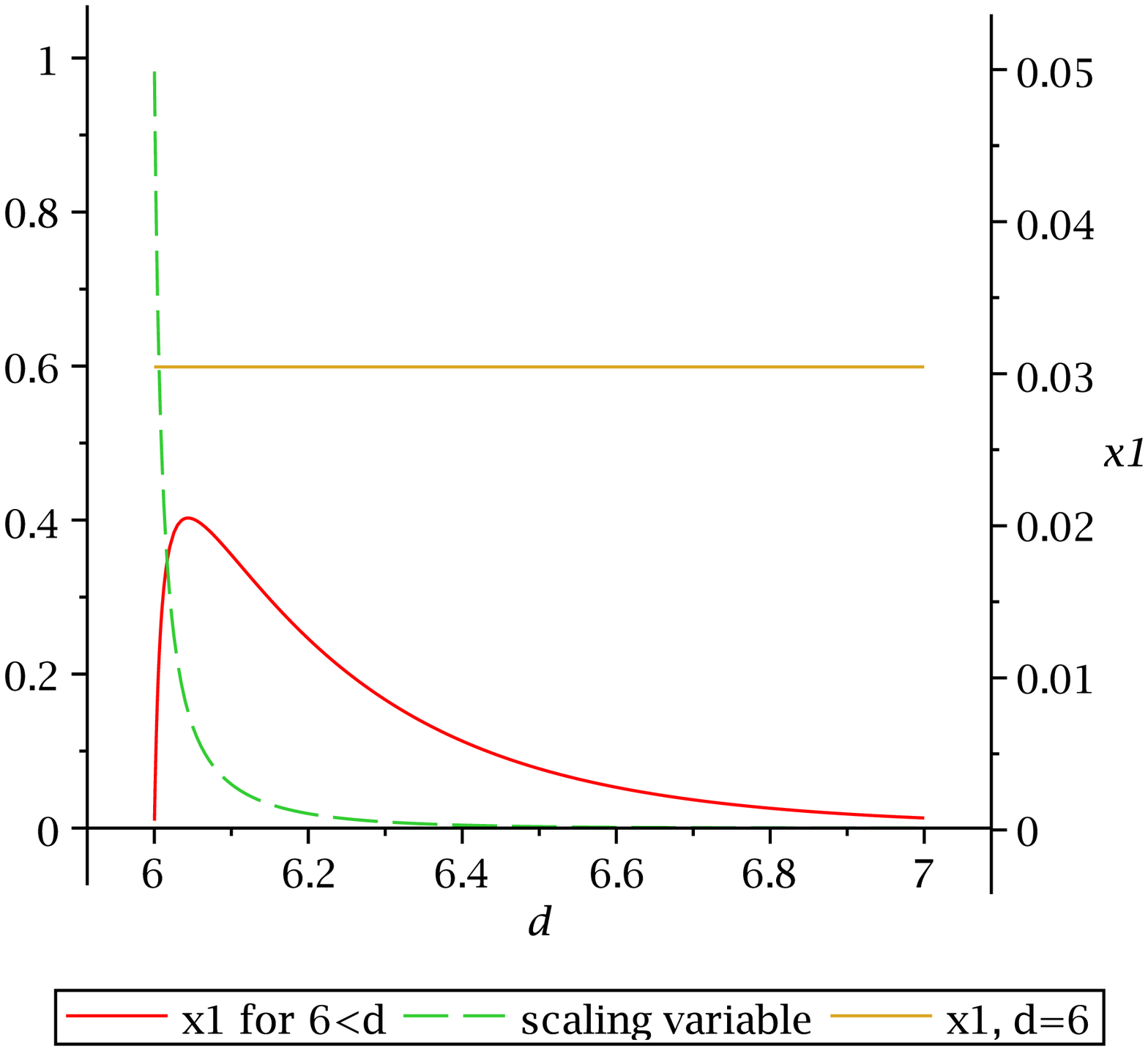}%
%}
\vspace{20pt}
%\hspace{190pt}\quad(a)\qquad\hspace{170pt}(b)\hspace{190pt}
\begin{center}
(a)\hspace{200pt}(b)
\end{center}
\end{figure}

For a thorough analysis of the $d$-dependence of $x_1$ while crossing the
upper critical dimension, we recollect here the one-loop truncated results from
previous sections; see Eqs.\ (\ref{x1_d>6}), (\ref{x1_d=6}) and (\ref{x1_d<6}).
The goodness of these approximations depends on the smallness of the scaling
variable, which is defined and expressed in terms of the bare parameters $w^2$
and $\tau$ as follows:
\begin{equation}\label{cases}
\mathord{\text{scaling variable}}=
\begin{cases}
\frac{2}{|\epsilon|}\,\tilde w^2\,r^{|\epsilon|/2}=
\frac{2}{|\epsilon|}\,w^2\,\tau^{|\epsilon|/2}\,
\big(1+\frac{2}{|\epsilon|}w^2\big)^{-1-\frac{5}{6}|\epsilon|} & \quad\text{$d>6$,
see (\ref{solution}),}\\[10pt]
w^2\,\big(1-w^2\ln r\big)^{-1}=w^2\,\big(1+\frac{5}{3}w^2\ln w^2-w^2\ln \tau\big)^{-1}
& \quad\text{$d=6$, see (\ref{tau_vs_r}),}\\[10pt]
\frac{2}{\epsilon}\,\tilde g\,r^{\epsilon/2}=\tau^{w^{*2}}\,
\Big(\frac{w^{*2}}{w^2}\Big)^{\frac{5}{3}{w^{*2}}}\,\Big(1-\frac{w^{*2}}{w^2}\Big),
\quad w^{*2}=\frac{\epsilon}{2}
& \quad\text{$d<6$, see (\ref{r_g}).} 
\end{cases}
\end{equation}
This scaling variable is displayed --- for a chosen pair of bare values
$w^2=0.005$ and $\tau=0.0001$, both much smaller than one, as it should be
in this approximation --- as a function of dimension
$d$ below [Fig.\ (\ref{scaling_variable}a)] and above [Fig.\ (\ref{scaling_variable}b)]
6, where it takes $\approx 0.005$. $x_1$ is also shown in this figure, with the
awkward behaviour of approaching zero from both sides of the upper critical
dimension six, while $x_1\approx 0.03$ in $d=6$. It is clear, however, from the
figure that our approximation breaks down when approaching $d=6$ from either side,
as the scaling variable goes to unity in that limit. As a matter of fact, it must be
stipulated that the scaling variable be at least as good as in $d=6$, i.e.\ 
$\lesssim  0.005$. Therefore, the range of applicability of our approximation
(for the given $w$ and $\tau$)
is constrained to $d\approx 5.99$ and $d\gtrsim 6.4$, respectively. (Note that the chosen
$w$ is just the fixed point when $d=5.99$.) Representative values of $x_1$ in these ranges,
together with the six-dimensional case, are presented in Tab.\ \ref{table}. It can be concluded
from this example that $x_1$ keeps on being monotonically increasing when lowering dimensions
through 6. Nevertheless, a discontinuity of $x_1(w^2,\tau)$ at $d=6$ cannot be excluded. An
extrapolation of the data from the range $d\gtrsim 6.4$, using an exponential and/or a power law
fit, provides $x_1(0.005,0.0001)\approx 0.026-0.028$, a value somewhat lower
%consistent with the six-dimensional
than the six-dimensional 
one, $0.030$, when considering the scaling variable as a measure of the relative error (it is $\approx 0.005$
in six dimensions, see Table \ref{table}). A similar extrapolation from the $d<6$ side,
however, does not exist.
\begin{table}
\caption{$x_1$ around six dimensions shows monotonically increasing behaviour
with decreasing dimensionality. The smallness of the scaling variable verifies the approximation.
The bare parameters $w^2=0.005$ and $\tau=0.0001$ are the same as in Fig.\ \ref{scaling_variable}.}
\label{table}
\vspace{20pt}
\begin{tabular}{|c|c|c|}
\hline
$d$  &  $\quad x_1\times 10^2\quad$  & \quad scaling variable \quad\\
\hline
6.8  & \quad 0.1287 \quad       & 0.000308\\
\hline
6.6  &\quad 0.2648 \quad       & 0.001026\\
\hline
6.4 & \quad 0.5649 \quad        & 0.003834\\ 
\hline
6   & \quad 2.9943 \quad         & 0.004991\\
\hline
\quad 5.99005\quad & \quad 2.9993\quad & 0.004776\\
\hline
5.99 & \quad 3.0000 \quad & 0\\
\hline
\quad 5.98995\quad &\quad 3.0006\quad & -0.004774\\
\hline
\end{tabular}
\end{table}

\begin{figure}
\caption{$x_1$ as a function of the reduced temperature $\tau$; $w^2=0.005$.
(a): $d\le 6$ and (b): $d\ge 6$. $x_1$ is zero at criticality, i.e.\ for $\tau=0$,
when $d\ge 6$. On the contrary, it is nonzero for $d<6$ and has the universal value
$x_1=3\,\epsilon+O(\epsilon^2)$ at $T_c$. }
\label{x1_vs_tau}
\vspace{20pt}
\includegraphics[scale=0.3]{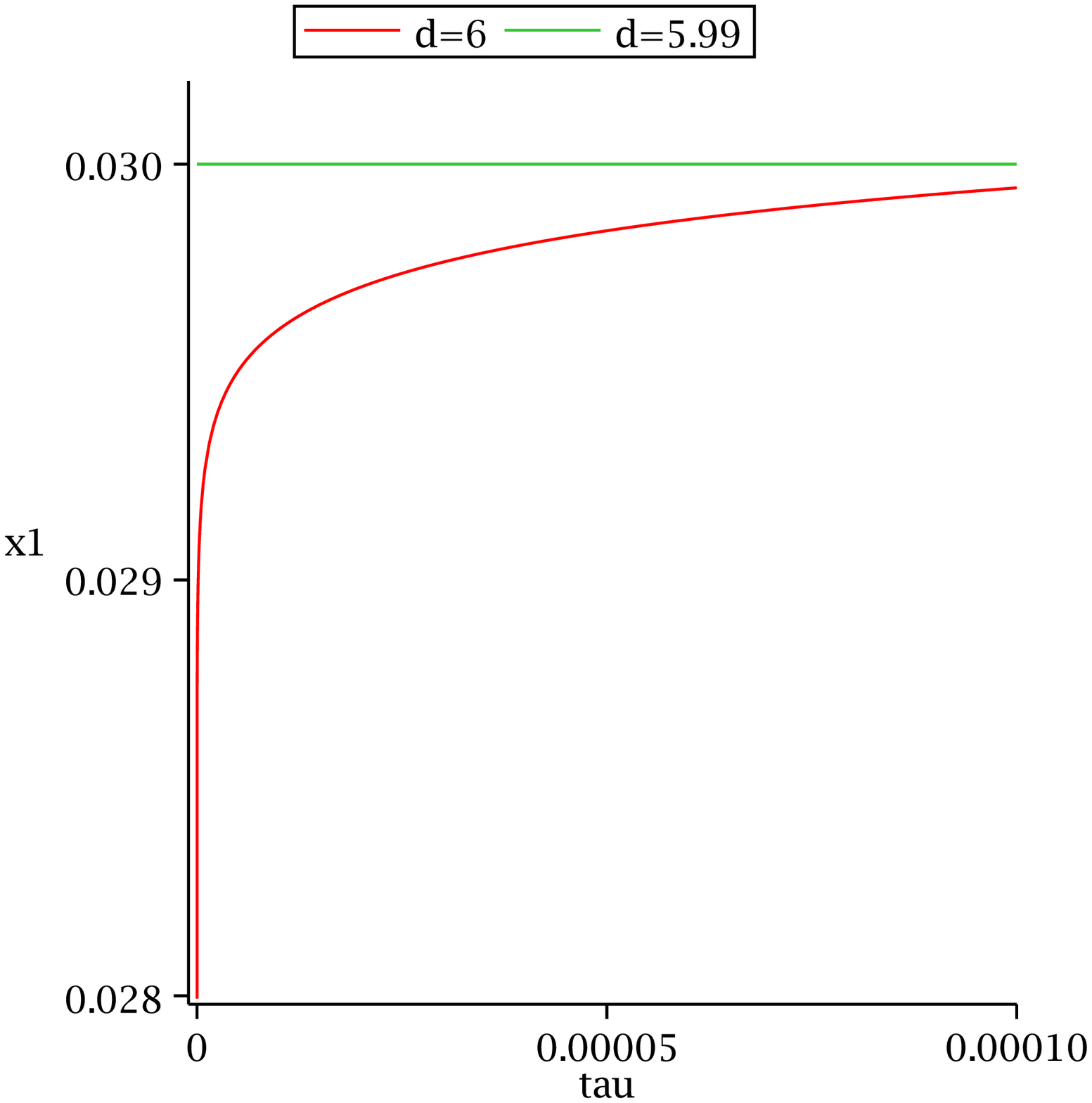}
\qquad\qquad\qquad
\includegraphics[scale=0.3]{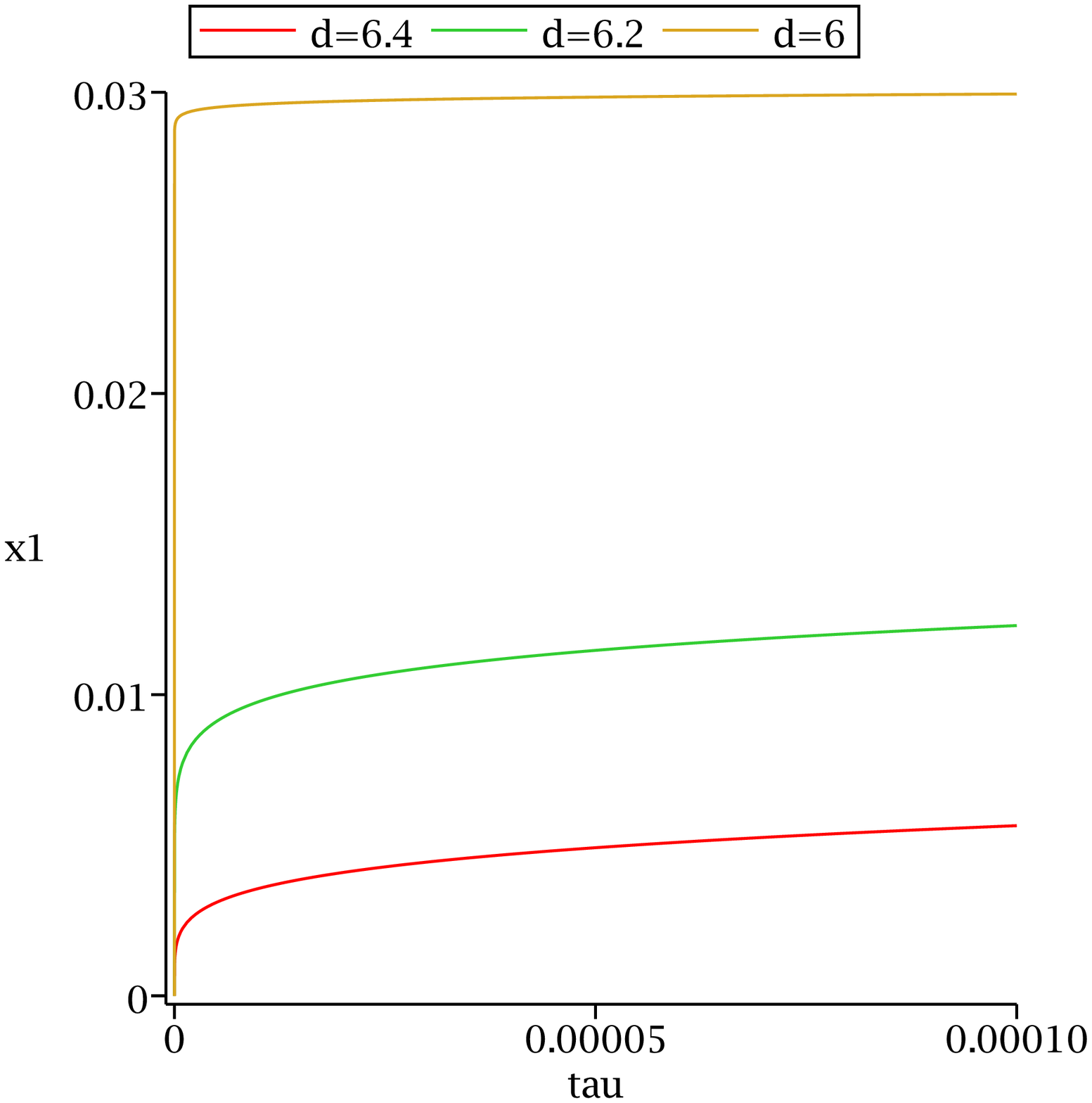}
\begin{center}
(a)\hspace{240pt}(b)
\end{center}
\vspace{20pt}
\end{figure}

Below six dimensions $x_1$ has only a slight temperature dependence, and it becomes nonzero
and universal at criticality:
\[
x_1=[3\,\epsilon+O(\epsilon^2)]+C\,\tau^{\frac{\epsilon}{2}+\dots}+\dots\,,\qquad d=6-\epsilon\,;
\]
$C$ is a nonuniversal, i.e.\ $w$-dependent, amplitude. The typical behaviour for both below and
above six dimensions is displayed in Fig.\ \ref{x1_vs_tau}, the value of the cubic coupling is
kept $w^2=0.005$. The tendency of an increasing $x_1$ while lowering the dimension is again obvious.
The vertical scale was magnified in the left subfigure (a) to show the qualitative difference
between the 6- and 5.99-dimensional curves. 

The critical field along the AT-line, for a given pair of bare parameters
$w^2$ and $\tau$, can be analysed using results from previous sections. See Eqs.\ 
(\ref{GammaR4}), (\ref{AT_d>6}) for $d>6$, and (\ref{AT_d=6}) for $d=6$.
Below six dimensions, if we wish to move somewhat away from the fixed point, the zero
of the expanded equation (\ref{hat_GammaR}) must be found, providing [instead of
(\ref{AT_d<6})]: 
\[
\tilde{h^2}=4\,w^*r^{\delta\,\beta}\,\big(1+8\,\frac{\tilde g}{\epsilon}\,r^{\epsilon/2}\big)\,
,\qquad \frac{\tilde g}{\epsilon}\,
r^{\epsilon/2}\ll 1\,\quad\text{and}\,\quad 0<\epsilon\ll 1\,.
\]
Eqs.\ (\ref{r_g}), (\ref{h}) and (\ref{cases}), together with $\delta\beta=2+\frac{3}{2}\epsilon$, give
us the critical field as
\[
w\,h^2=4\,w^2\,\tau^{2+3\,w^{*2}}\,\Big(\frac{w^{*2}}{w^2}\Big)^{4+5\,w^{*2}}\,
\bigg[1+4\,\tau^{w^{*2}}\,
\Big(\frac{w^{*2}}{w^2}\Big)^{\frac{5}{3}{w^{*2}}}\,\Big(1-\frac{w^{*2}}{w^2}\Big)\bigg],
\quad w^{*2}=\frac{\epsilon}{2}\,.
\]

\begin{figure}
\caption{Almeida-Thouless line ($wh^2$ versus $\tau$)
of the field theoretic model with $w^2=0.005$ for three different
dimensions.}
\label{AT_figure}
\vspace{20pt}
\includegraphics[scale=0.3]{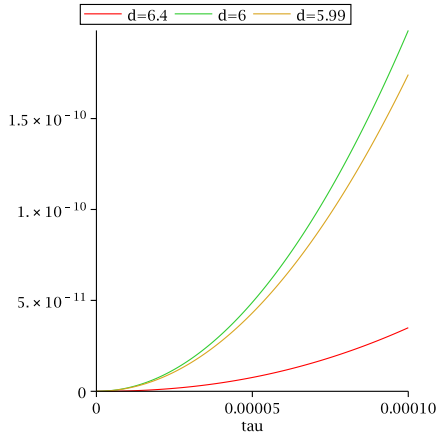}
\vspace{20pt}
\end{figure}

The critical field where RSB sets in as a function of temperature (i.e.\ 
the AT line) --- or more precisely $wh^2$ as a function of $\tau$ ---
is shown in Fig.\ \ref{AT_figure} for three different dimensions at fixed
cubic coupling $w^2=0.005$. The curve for $d=5.99$ (note that the system is at exactly the fixed
point then) is significantly below the six-dimensional one. It is easy to see that this follows
directly from the exponent inequality $\delta\beta-2=\frac{3}{2}\epsilon+\dots>0$. To see clearly the
behaviour of the critical field above and below six dimensions for decreasing $d$, it is tabulated
in Table \ref{AT_table} for the system with $w^2=0.005$ and $\tau=0.0001$.
The last three rows of this table show that the kind of monotonicity found above six dimensions
is restored below it, i.e.\ the critical field increases with decreasing dimensions. It must be remarked,
however, that around the last dimension value $d=5.98995$, the error%
\footnote{The relative error can be estimated as being proportional to the square of the
scaling variable.}
of our approximation
starts to be the same order of magnitude as the variation of the critical field itself. The range
where this one-loop perturbative method is applicable below the upper critical dimension
is certainly very narrow.
\begin{table}
\caption{Critical field values around six dimensions for $w^2=0.005$ and
$\tau=0.0001$. Below the critical field replica symmetry is broken. The scaling variable's
values are, of course, the same as in Table \ref{table}.}
\label{AT_table}
\vspace{20pt}
\begin{tabular}{|c|c|c|}
\hline
$d$  &  $\quad wh^2\times 10^{10}\quad$  &\quad scaling variable \quad\\
\hline
6.8   & \quad 0.0827 \quad & 0.000308 \\
\hline
6.6   & \quad 0.1680 \quad & 0.001026 \\
\hline
6.4  &  \quad 0.3497 \quad & 0.003834 \\
\hline
6    &  \quad 1.9849 \quad & 0.004991 \\
\hline
5.99005 &  \quad 1.7410 \quad & 0.004776 \\
\hline
5.99    &  \quad 1.7419 \quad & 0   \\ 
\hline
5.98995 & \quad  1.7421 \quad & -0.004774 \\
\hline
\end{tabular}
\end{table}

As a conclusion, we can confidently claim that RSB survives below six dimensions
in the cubic replica field theory
representing the Ising spin glass. We focused on two quantities which are strongly related to RSB:
the breakpoint of the order parameter function $x_1$ and the Almeida-Thouless line. A combination of
the perturbative one-loop method with a simple two-parameter renormalization group (which is 
correct near the critical fixed point) provided reliable results in all the three ranges of
dimensionalities, i.e.\ for $d$ larger, equal, and less than six. The calculations above and below six
dimensions are rather different, due to the Gaussian versus nontrivial fixed point governing critical
behaviour in the two cases. The applied perturbative method makes it impossible
to approach closely the upper critical dimension: the range of dimensions where the approximation is
correct for a given system (i.e.\ for given $w$ and $\tau$) is very narrow and close to 6 when
$d<6$, whereas it is $d\gtrsim 6.2$ when $d>6$ (and the farther
we are from $d=6$, the better the approximation).
The six-dimensional case needs special care along the way systems at their upper
critical dimension are commonly studied \cite{PfeutyToulouse}. The logarithmic temperature dependences
obtained are quite similar to those in ordinary systems at their upper critical dimension.

Above six dimensions both $x_1$ and the critical field are monotonically increasing for decreasing $d$,
and this tendency persists for $d<6$. There seems to be, however, a discontinuity of the critical field
at $d=6^-$: the AT line for $d\lessapprox 6$ is significantly below the six-dimensional one,
see Fig.\ \ref{AT_figure} and Table \ref{AT_table}. Nevertheless, we can notice that the trend of
increasing dominance of RSB for decreasing space dimensions persists even below six dimensions.

As a final remark, we recall that for $d<6$, $x_1$ gains the qualitatively new feature of
being nonzero (and universal!) at criticality. This might suggest a kind of first order transition.
That this is not the case can be clearly seen by displaying the order parameter function using
Eqs.\ (\ref{q(x)}) and (\ref{q_1}):
\[
q(x)=q_1\,\hat q(x/x_1)\,,\qquad \text{where} \qquad q_1\sim q\sim \tau^{\beta}\,.
\]  
An elaboration of the equation of state along the lines of Sec.\ \ref{RSB_vs_RS} for $d<6$
(which is out of the scope of the present paper, and is left for a future publication),
proves that, next to the spin glass transition, $\hat q$ is a function independent of temperature,%
\footnote{This has been suggested in Ref.\ \cite{beyond}, see Eq.\ (155) of it.}
and thus nontrivial even at criticality. The prefactor $q_1$, however, disappears at $T_c$ ensuring
continuity of the order parameter through the spin glass transition.

\begin{acknowledgments}
We are extremely grateful to Imre Kondor for his thorough review of the paper prior to publication,
and also for his useful suggestions. 
\end{acknowledgments}

%\bibliography{spinglass}
%\begin{thebibliography}{16}
%\end{thebibliography}

\end{document}